\documentclass[prd,11pt,a4paper,nofootinbib,showpacs,twoside,tightenlines]{revtex4}

\usepackage[dvips]{color}

\usepackage{graphicx,mathrsfs,amsmath,amssymb,amsthm,amsopn,mathbbol,bm}

\newenvironment{Equation*}{\equation}%
{\endequation}
\newenvironment{Eqnarray*}{\eqnarray}%
{\endeqnarray}

\newcommand{\TrR}{\Tr_{\R}}
\newcommand{\TrT}{\Tr_{\text{Th}}}
\newcommand{\CT}{\mathcal{C}_{\text{Th}}}
\newcommand{\CR}{\mathcal{C}_{\R}}
\newcommand{\funcint}[2]{\int_{\mathcal{C}} \text{d}({#2}) \; #1}

\newcommand{\R}{\ensuremath{\mathbb{R}}}
\newcommand{\ii}{\ensuremath{\mathrm{i}}}
\newcommand{\op}[1]{\ensuremath{\bm{\mathrm{#1}}}}
\newcommand{\erw}[1]{\ensuremath { %
    \left \langle {#1} \right \rangle}}
\newcommand{\D}{\ensuremath{\mathrm{D}}}
\renewcommand{\d}{\ensuremath{\mathrm{d}}}
\newcommand{\funcd}[2]{\frac{\delta #1}{\delta #2}}
\DeclareMathOperator{\Tr}{Tr}
\newcommand{\Lag}{\ensuremath{\mathscr{L}}}
\newcommand{\feynint}[1]{\ensuremath{\int \frac{\mathrm{d}^{d} {#1}}{(2
\pi)^{d}}}}
\newcommand{\fint}[1]{\ensuremath{\int \frac{\mathrm{d}^4
      #1}{(2\pi)^4}}}
\DeclareMathOperator{\sign}{sign}
\DeclareMathOperator{\im}{Im}
\newcommand{\comm}[2]{\ensuremath{ \left[ {#1}, {#2} \right]_{-} }}

\newcommand{\natop}[2]{\genfrac{}{}{0pt}{1}{#1}{#2}}
\begin{document}

\title{Renormalization in Self-Consistent Approximation schemes at \\
  Finite Temperature I: Theory} 
\author{Hendrik van Hees, J{\"o}rn
  Knoll} \affiliation{GSI Darmstadt} 
\date{August 29, 2001}

\begin{abstract}
  
  Within finite temperature field theory, we show that truncated
  non-perturbative self-consistent Dyson resummation schemes can be
  renormalized with local counter terms defined at the vacuum level.
  The requirements are that the underlying theory is renormalizable
  and that the self-consistent scheme follows Baym's $\Phi$-derivable
  concept.  The scheme generates both, the renormalized
  self-consistent equations of motion and the closed equations for the
  infinite set of counter terms.  At the same time the corresponding
  2PI-generating functional and the thermodynamical potential can be
  renormalized, in consistency with the equations of motion. This
  guarantees the standard $\Phi$-derivable properties like
  thermodynamic consistency and exact conservation laws also for the
  renormalized approximation schemes to hold. The proof uses the
  techniques of BPHZ-renormalization to cope with the explicit and the
  hidden overlapping vacuum divergences.

\end{abstract}


\maketitle

\section{Introduction}

In recent years the question how to appropriately treat particles in a
hot and dense medium has continuously gained growing interest in many
areas of physics ranging from plasma physics, condensed matter physics
to nuclear and particles physics. Within a Green's function formalism
a consistent treatment of such phenomena frequently leads to consider
dressed propagators, which follow from non-perturbative Dyson
resummation schemes, rather than perturbative ones, in particular, if
damping width effects play a significant role.

Already in the early sixties, based on a functional formulation of
Luttinger and Ward \cite{lw60} and Lee and Yang \cite{leeyang61},
Kadanoff and Baym \cite{bk61} considered a class of self-consistent
Dyson approximations.  Baym reformulated this in terms of a
variational principle, defining the so called $\Phi$-derivable
approximations \cite{baym62}. Since in principle the truncation of the
diagrammatic series of the functional $\Phi$ can be at arbitrary level
it gives rise to a variety of approximations including Hartree and
Hartree--Fock as the simplest schemes. The main virtue of this concept
is that the resulting equations of motion are conserving and the
corresponding equilibrium limit is thermodynamically consistent.  This
functional treatment constitutes the basis for the two-particle
irreducible (2PI) diagram technique, where the functional $\Phi$
generates the driving terms for the equations of motion, like the
self-energy. Later the concept was extended to the relativistic case
and formulated within the path integral approach by Cornwall, Jackiw
and Tomboulis \cite{cjt74}.  There is no formal problem to extend the
formalism to the Schwinger Keldysh real-time path method
\cite{Sch61,kel64} applicable to the general case of non-equilibrium
many-body theory.

Despite such early conceptual formulations most applications of
self-consistent approximations were pursued on the Hartree or
Hartree-Fock level sometimes supplemented by RPA resummations (see,
e.g., \cite{chin77,baymgrin}) or perturbative estimates of higher
order corrections.  Thus, essentially mean field corrections to the
self-energies were considered. Genuine two-point or even multi-point
contributions to the self-consistent self-energy, which give rise to a
finite damping width, imply a new level of complexity.  Various new
conceptual problems, like leaving the quasi-particle picture or the
issue of renormalization come in with considerable complications for
the numerical solutions of such problems. In the pioneering work of
Bielajew and Serot \cite{bielserot84} for the first time the
renormalization of self-consistent two-point self-energy loops were
investigated at zero temperature but finite matter density.

In recent years with the special interest in dense hadronic matter
problems the $\Phi$-derivable schemes with higher order self-energy
terms were used to derive transport equations
\cite{kv97,knoll98,ikv99-2,Leupold99} from the corresponding
Kadanoff-Baym equations \cite{kb61} for the consistent and also
conserving \cite{KIV01} transport treatment of particles with finite
spectral width beyond the quasi-particle approximation. Also first
investigations involving finite mass width effects on vector mesons
were investigated within a self-consistent Dyson resummation scheme
\cite{vHK2001}.  In most of these cases, however, the question of
renormalization was circumvented by taking into account the imaginary
part of the self-energy only, while the real part was neglected, or
cut-off recipes -- mostly symmetry violating -- were employed, or even
the counter terms were chosen temperature dependent!  Yet, especially
in the study of phase transitions, e.g., within chiral hadronic models
of QCD, or for non-perturbative corrections of hard thermal loop
approaches to QCD, e.g., within a $\Phi$-derivable scheme
\cite{ianc00,pesh00} it is important to consistently take into account
both, real and imaginary parts of the self-energies.

Therefore in this paper we address the more formal question of
renormalizability of such non-perturbative approximations.  We
essentially concentrate on the thermodynamic equilibrium case and show
how to obtain finite self-consistent dynamical quantities like the
in-medium equations of motion and the self-energy of the particles and
thermodynamic quantities like the pressure and the entropy. For
definiteness and clarity of the presentation we use the
$\phi^4$-theory as the most simple example to study the related
questions. The results and techniques can easily be transfered to
other theories.

The paper is organized as follows: In section \ref{sect2} we briefly
summarize Baym's $\Phi$-functional using the combined real and
imaginary time contour appropriate for thermal equilibrium within the
path integral formalism \cite{cjt74}.

In section \ref{sect3} we derive the general formalism for the
renormalization of the self-consistent self-energy at finite
temperature and the in-matter generating functional $\Gamma$. With the
help of Weinberg's convergence theorem \cite{wein60} and the
BPHZ-formalism of renormalization theory \cite{bp57,Zim69} we show
that, in close analogy to perturbative renormalization (see, e.g.,
\cite{col86,kap89,lebel}) \emph{any $\Phi$-derivable self-consistent
  approximation scheme can be rendered finite by subtracting pure
  vacuum counter terms} given by closed recursive equations.  Indeed
the main complication arises from the fact that the self-consistent
propagator is involved in divergent loops which gives rise to
``hidden'' divergences which have to be resolved.  This leads to a
Bethe-Salpeter equation for the divergent vacuum pieces with a kernel
compatible with the functional $\Phi$, which needs to be renormalized.
The renormalized equations of motion for the self-consistent
propagator are shown to be consistent with the renormalized 2PI
generating functional which proves the consistency of counter terms at
both levels. The diagrammatical interpretation shows that in strict
analogy to perturbative renormalization of thermal quantum field
theory this procedure can be interpreted as renormalization of the
wave functions, the mass and the coupling constants \emph{in the
  vacuum}. We also give a closed expression for the renormalized
self-consistent thermodynamical potential.

Numerical solutions for the renormalized self-consistent Dyson
equations beyond the standard Hartree approximation up to the
self-consistent sunset-diagram level could be achieved; the results
are discussed in the second paper of this series \cite{vHK2001-Ren-II}.

\section{$\Phi$-derivable approximations}
\label{sect2}

\begin{figure}[t]
\centering{\includegraphics{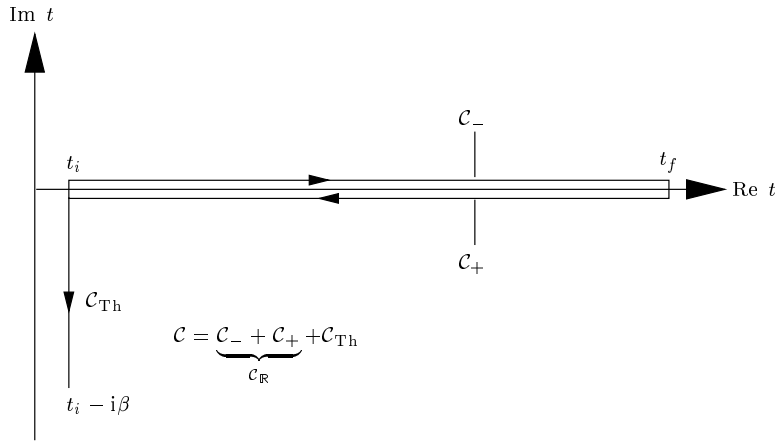}}
\caption{The Schwinger-Keldysh real-time contour modified for the
  application to thermal equilibrium of quantum field theory.}
\label{figure1}
\end{figure}

In the case of thermal equilibrium the real and imaginary time
formalism can be combined by extending the Schwinger-Keldysh contour
$\CR$, running from $t_i$ to $t_f$ and back to $t_i$, by appending a
vertical part $\CT$ running from $t_i$ to $-\ii \beta$ (see fig.
\ref{figure1}). One uses the fact that the factor $\exp(-\beta
\op{H})$ in the canonical density operator can be formally treated as
a time evolution in imaginary time direction. The functional integral
for bosonic fields has to be taken over all fields fulfilling the
\emph{periodic boundary condition} $\phi(t_i-\ii \beta)=\phi(t_i)$
which leads to the Kubo-Martin-Schwinger (KMS) condition for the
Green's functions \cite{lvw87}. Since the equilibrium state is
invariant in time one can take $t_i \rightarrow -\infty$ and $t_f
\rightarrow +\infty$ which is convenient when formulating the theory
in energy-momentum space via a Fourier transformation.

In addition to the usually introduced \emph{one-point auxiliary
  external source} also a \emph{two-point auxiliary external source}
is included. Variations of the latter generate contour-ordered
expectation values of the form $\erw{\mathcal{T}_{\mathcal{C}}
  \op{\phi}(x_{1}) \op{\phi}(x_{2})}$ where $\mathcal{C}$ denotes the
extended Schwinger-Keldysh time contour, and
$\mathcal{T}_{\mathcal{C}}$ stands for the ordering of the operators
due to the ordering of the time arguments along this contour.

The corresponding generating functional is defined within the
path integral formalism of quantum field theory as
\begin{equation}
\label{1}
Z[J,B]=N \int \D \phi \; \exp \left[\ii S[\phi] + \ii \funcint{{J_1}
      \phi_1}{1}  +
 \frac{\ii}{2} \funcint{B_{12} \phi_1 \phi_2}{12} \right], 
\end{equation}
where $N$ is a normalization constant chosen below. Here and in the
following the shorthand notation
\begin{equation} 
\label{2}
\funcint{f_{12\ldots n}}{12\ldots n}=\int_{\mathcal{C} \times
  \R^{d-1}} \d^{d} x_1 \cdots \d^{d} x_n f(x_1,\ldots,x_n)
\end{equation}
for integrations in $d$-dimensional space is used (in the sense of
dimensional regularization\footnote{We use this notation for
  convenient regularization only, in order to write down sensible
  non-renormalized functionals.}). The time integration has to be
performed along the time contour introduced above. It is clear that
also the action functional $S$ has to be defined as the $\mathcal{C}
\times \R^{d-1}$-integral of the Lagrangian.

The generating functional $W$ for \emph{connected diagrams} reads
\begin{equation}
\label{3}
W[J,B]=-\ii \ln (N Z[J,B]).
\end{equation}
The \emph{mean field} and the \emph{connected Green's functions}
are defined by
\begin{equation}
\label{4}
{\varphi_1} =\funcd{W}{J_1}, \quad {G_{12}}=-\frac{\delta^2
  W}{\delta J_1 \delta J_2} \quad \Rightarrow \quad \funcd{W}{B_{12}} =
\frac{1}{2} (\varphi_1 \varphi_2 + \ii G_{12}).
\end{equation}
The last formula immediately follows from the definition of the
partition sum (\ref{1}) and (\ref{3}) using the Feynman-Kac
formula
\begin{equation}
\label{5}
\erw{\mathcal{T}_{\mathcal{C}} \op{\phi}(x_{1}) \cdots
  \op{\phi}(x_{n})} =
\int \D \phi \; \phi(x_{1}) \cdots \phi(x_{n}) \exp(\ii S[\phi]).
\end{equation}
Since the real-time part of this contour is closed by itself it can
be shown that the functional (\ref{1}) factorizes into the real-time
part and the imaginary time part. Thus the Feynman rules for the
calculation of connected Green's functions apply separately to the
real and the imaginary part of the contour \cite{gel94}, since the
functional (\ref{3}) splits into a sum of the contributions from the
vertical and the real-time part of the contour respectively:
\begin{equation}
\label{5b}
Z[J,B]=Z_{\R}[J,B] Z_{\text{Th}}[J,B], \;
W[J,B]=W_{\R}[J,B] + W_{\text{Th}}[J,B]-\ii \ln N.
\end{equation} 

Within the real-time part the effect of the heat bath is completely
taken into account by means of the analytical properties of the
Green's function. It is uniquely determined by the KMS-condition which
itself is a consequence of the above mentioned periodic boundary
conditions for bosons within the path integral (see appendix
\ref{appa} for details about the analytic properties of Green's
functions and self-energies).

By a functional Legendre transformation in $\varphi$ and $G$ one
obtains the \emph{effective quantum action}:
\begin{equation}
\label{6}
\Gamma[\varphi,G]=W[J,B]-\funcint{\varphi_1 J_1}{1} - \frac{1}{2}
\funcint{(\varphi_1 \varphi_2 + \ii G_{12})B_{12}}{12}-\ii \ln N.
\end{equation} 
Now, as is well known from the usual functional formalism of quantum
field theory, a formal saddle point expansion of the effective quantum
action is an expansion in orders of $\hbar$ around the classical
solution, where $G$ is considered as an independent quantity, gives
\begin{equation}
\label{7}
\Gamma[\varphi,G] = S[\varphi] + \frac{\ii}{2} \Tr \ln (M^2 G^{-1}) +
\frac{\ii}{2} \funcint{D_{12}^{-1} 
  (G_{12}-D_{12})}{12} + {\Phi[\varphi,G]}.
\end{equation}
Herein the free propagator in the presence of a mean field is given by
\begin{equation}
\label{8}
 D_{12}^{-1}=\frac{\delta^2 S[\varphi]}{\delta
  \varphi_1 \delta \varphi_2}.
\end{equation}
The arbitrary constant $M^2$ account for the overall normalization
and cancels for any physical quantities as we shall see below. 

In the case of an ideal gas it is sufficient to subtract the pure
vacuum part to render this functional finite, which leads to the well
known result. At $T=0$ this subtraction corresponds to the
renormalization of the total ground-state energy to zero.

As we shall discuss below in the case of interacting particles this
description is not sufficient to render the effective action and thus
the pressure finite, since we need additional subtractions of vacuum
sub-divergences to renormalize it.

In the above sense the functional $\Phi$ in (\ref{7}) contains the parts
of order $O(\hbar^2)$ and higher. Since simple power counting shows
the $\hbar$-order of diagrams to be identical with the number of
loops, $\Phi$ as a functional of $G$ consists of all closed diagrams
with at least $2$ loops.  The lines within the diagrams stand for
dressed Green's functions $G$ while the vertices are the bare vertices
of the classical action with presence of the background field
$\varphi$ which can be immediately read off from $S_I[\varphi+\phi']$
around $\phi'=0$ beginning at order $\phi'{}^3$.

The equations of motion are now given by the fact that we like to
study the theory with vanishing auxiliary sources $j$ and $B$. From
(\ref{4}) and (\ref{6}) we obtain
\begin{equation}
\label{8b}
\funcd{\Gamma}{\varphi_1}=-j_{1} \stackrel{!}{=}0, \;
\funcd{\Gamma}{G_{12}} = -\frac{\ii}{2} B_{12} \stackrel{!}{=}0.
\end{equation}
Using (\ref{7}) the equations of motion read
\begin{equation}
\label{8c}
\begin{split}
\funcd{S}{\varphi_1} & =-\frac{\ii}{2}
\funcint{\funcd{D_{1'2'}^{-1}}{\varphi_1} G_{1'2'}}{1'2'}
-\funcd{\Phi}{\varphi_1} \\
\Sigma_{12} &:=D_{12}^{-1}-G_{12}^{-1}=2 \ii \funcd{\Phi}{G_{12}}.
\end{split}  
\end{equation}
The first line is of the form of a Klein-Gordon equation with the
quantum corrections to the classical theory on the right hand side.
The second equation is the \emph{Dyson equation} and shows that the
variation of the $\Phi$-functional with respect to $G$ is the
self-energy defined with respect to the classical field dependent
propagator (\ref{8}). This shows that the $\Phi$-functional must be
\emph{two-particle irreducible}. No propagator line must contain a
self-energy insertion. In other words the closed diagrams representing
contributions to $\Phi$ must not split in disconnected pieces when
cutting two lines. Diagrammatically the derivative of a functional
with respect to $G$ corresponds to opening one line of the diagrams
representing it. In that sense $\Phi$ is the generating functional for
\emph{skeleton diagrams for the self-energy} where the lines represent
fully dressed propagators. Thus the functional formalism avoids double
counting in a natural way by omitting all non-skeleton diagrams from
the Dyson-resummed equations of motion. Altogether the
$\Phi$-functional formalism provides a closed system of exact
equations of motion for the full $2$-point function and the full mean
field. Solving these equations would be equivalent to finding the full
propagator of the quantum field theory which of course is impossible
in practice.

One obtains approximations by truncating the series for $\Phi$ at a
certain vertex or loop order (which corresponds to the respective
order in the coupling $\lambda$ or $\hbar$ respectively), while
preserving the forms (\ref{8b}) and (\ref{8c}) of the self-consistent
equations of motion.  Approximations of this kind respect the
conservation laws for the expectation values of Noether currents for
symmetries which are linearly operating on the field operators
(including space-time symmetries and the according conserved
quantities as energy, momentum and angular momentum)
\cite{baym62,kv97}.

In the case of thermal equilibrium, setting the mean field and the
propagator to the solution of the self-consistent equations the
effective action gives the grand canonical potential $\Omega=-T
\Gamma[\varphi,G]_{J,K=0}=-T \ln Z(\beta)$ \cite{lvw87}. Since the
real-time part of the contour in figure \ref{figure1} is closed it
vanishes for the solution of the equations of motion. A short summary
about the analytic continuation from the real to the imaginary time
formalism is given in appendix \ref{appa}.

Thus the formalism leads to a well defined treatment for bulk
thermodynamical quantities of the system (like energy, pressure,
entropy, etc.).

All these quantities can be calculated either with real-time Green's
functions or with the corresponding imaginary time functions, because
as summarized in appendix \ref{appa} real and imaginary-time
propagators are connected by the analytic properties of the Green's
functions originating from the KMS-condition. For our purpose the
real-time formalism is preferred, because of its simplicity with
respect to the analytic structure of Green's functions, which
  easily permits to deal with the mixture of finite temperature and
  vacuum pieces occurring in the subtraction scheme.  This also
avoids the necessity to perform an analytic continuation from
imaginary time to real-time Green's functions which is complicated to
obtain for numerical results.

In order to exemplify the method we apply the formalism to
$\phi^4$-theory with the Lagrangian
\begin{equation}
\label{9}
\Lag=\frac{1}{2}(\partial_{\mu} \phi) (\partial^{\mu} \phi) -
\frac{m^2}{2} \phi^2 - \frac{\lambda}{4!} \phi^4.
\end{equation}

\section{Renormalizability of $\Phi$-derivable approximations}
\label{sect3}

In this section we show that in close analogy to the renormalization
of perturbative diagrams also any $\Phi$-derivable self-consistent
approximation can be renormalized with help of local
temperature-independent counter terms.

The proof uses the same line of arguments as in the perturbative case:
The reason is, that the renormalization theory completely rests on
Weinberg's power counting theorem \cite{wein60} which is formulated
for a general class of Green's functions with a given asymptotic
behavior. It does not depend on the special form of the propagators.

The first step is a simple topological argument leading to the
superficial degree of divergence for a given diagram $\gamma$, which
for the $\phi^4$ theory simply follows from the number $E$ of external
lines \cite{itz80}
\begin{equation}
\label{sdod}
\delta(\gamma)=4-E.
\end{equation}
Due to field reflection symmetry only diagrams with an even number of
external lines are different from $0$, the only divergent functions
are those represented by diagrams where the number of external legs is
$0$ (i.e., contributions to the total mean energy and the
thermodynamical potential), $2$ (self-energy (Green's function)) and
$4$ (four-point vertex functions).

The second step is an expansion of the regularized un-renormalized
self-consistent self-energy around the self-consistent vacuum
propagator which shows that the asymptotic behavior of the diagrams
and sub-diagrams is ruled by their pure vacuum parts.

Then an equation of motion for the temperature-dependent ``infinite
part'' of the regularized self-energy is derived and it is shown that
it can be renormalized by a temperature independent subtraction
procedure.

\subsection{BPHZ-Scheme for the vacuum}

We first apply the BPHZ renormalization theorem
\cite{bp57,Zim69,zim70} for the vacuum. The only difference to the
perturbative case is that we apply it to diagrams with self-consistent
propagator lines. This is justified since Weinberg's power counting
theorem is independent of the special form of propagators but only
needs their asymptotic behavior stated above.

We summarize the BPHZ scheme as follows. A sub-diagram $\gamma$ of a
diagram $\Gamma$ is defined as any set of lines and vertices contained
in $\Gamma$ which itself builds a proper vertex diagram: $\gamma
\subseteq \Gamma$. A sub-diagram $\gamma$ is called
\emph{renormalization part} if its superficial degree of divergence
(or its dimension) is greater than or equal to $0$. In our case this
means it has at most four external legs. Two sub-diagrams $\gamma_1$
and $\gamma_2$ are called \emph{nested} $\gamma_1 \subseteq \gamma_2$
if $\gamma_1$ is a sub-diagram of $\gamma_2$. If they have no line or
vertex in common, $\gamma_1 \cap \gamma_2 =\emptyset$, they are called
\emph{disjoined}. If they are neither nested nor disjoined they are
called \emph{overlapping}: $\gamma_1 \circ \gamma_2$.

To any diagram $\Gamma$ we denote the integrand following from the
Feynman rules with $I_{\Gamma}$. For a set of pairwise disjoined
sub-diagrams $\gamma_1,\gamma_2,\ldots,\gamma_n$ we write the
integrand in terms of the integrands of the sub-diagrams
$I_{\gamma_j}$ and the remainder of the integrand denoted by
$I_{\Gamma \setminus \{\gamma_1,\ldots,\gamma_n \} }$, usually called
the \emph{reduced diagram}:
\begin{equation}
I_{\Gamma}=I_{\Gamma \setminus \{\gamma_1,\ldots,\gamma_n \} }
\prod_{j=1}^{n} I_{\gamma_j}.
\end{equation}
The original scheme by Bogoliubov and Parasiuk \cite{bp57} defines
recursively the integrand $R_{\Gamma}$ of the renormalized diagram. If
a diagram does not contain any renormalization part but is itself
divergent it is called \emph{primitively} divergent. In that case the
renormalized integrand is defined by $R_{\Gamma}=(1-t_{\Gamma})
I_{\Gamma}$. Herein $t_{\Gamma}$ is the operator of the Taylor
expansion with respect to the external momenta around $0$ up to the
order of the dimension $\delta(\gamma)$ of the divergent diagram,
which is in our case $4-E$:
\begin{equation}
\label{def-t}
t_{\gamma}
I_{\gamma}(p_{1},\ldots,p_{k}):=
\left\{
\begin{array}{l}\displaystyle
\sum_{j=0}^{\delta(\gamma)}
\frac{1}{j!} \sum_{\natop{\mu_{1},\ldots, \mu_{k}\ge
    0}{\mu_1+\ldots+\mu_k=j}} 
\left .
\frac{\partial^{j}I_{\gamma}(p_{1},\ldots,p_{k})}
{\partial   p_{1}^{\mu_{1}} \cdots   \partial p_{k}^{\mu_{k}}}
\right |_{p_{1}=\ldots=p_{k}=0}
p_{1}^{\mu_{1}}  \cdots p_{k}^{\mu_{k}}\\
0\quad\mbox{for}\quad \delta(\gamma)<0
\end{array}\right. 
\end{equation}
If the diagram is convergent the integrand is unchanged under
renormalization.

If the diagram is not only primitively divergent but contains
divergent sub-diagrams the integrand for the diagram with all
subdivergences subtracted is called $\bar{R}_{\Gamma}$ and the
renormalized integrand is defined by
\begin{equation}
R_{\Gamma}=\begin{cases}
\bar{R}_{\Gamma} & \text{if $\delta(\Gamma)<0$} \\
(1-t_{\Gamma})\bar{R}_{\Gamma} & \text{if $\delta(\Gamma) \geq 0$}.
\end{cases}
\end{equation}
From Weinberg's power counting theorem it follows that after this
recursive procedure the integral over the internal momenta of
$R_{\Gamma}$ is finite. The definition of the \emph{counter terms} by
the Taylor operator $t_{\gamma}$ for any renormalization part $\gamma$
of the diagram shows that these are polynomials in the external
momenta to the order $\delta(\gamma)$ and thus can be interpreted as
counter terms to the corresponding wave function normalization
factors, masses and coupling constants in the original Lagrangian.

Zimmermann solved Bogoliuobov's and Parasiuk's recursion with his
\emph{forest} formula. A forest is defined as any set of sub-diagrams
(including the empty set and the whole diagram itself) which are
\emph{pairwise non-overlapping}. One can depict these sets by drawing
boxes around the sub-diagrams and in a forest these boxes are not
allowed to overlap but they can be nested. A forest is restricted if
each of its boxes contains only renormalization parts. To each
restricted forest $\mathfrak{F}$ one associates again an integrand,
namely
\begin{equation}
\label{forest-ex}
\Omega_{\mathfrak{F}} = \tilde{\prod}_{\gamma \in \mathfrak{F}}
(-t_{\gamma}) I_{\Gamma} = \hspace{-3mm}\parbox{32mm}{\includegraphics{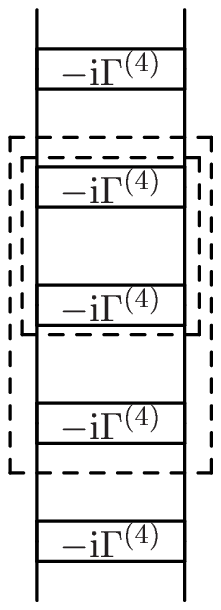}}
\end{equation}
The diagram to the right shows an example case for a typical ladder
diagram, which we shall consider in the following section. For this
case $\delta(\gamma)=0$ for all sub-diagrams and the diagram itself
such that for each box only the subdiagram value at vanishing external
momenta is to be subtracted. The tilde over the product sign in
(\ref{forest-ex}) stands for the fact that in case of nested diagrams
within the forest one has to apply the Taylor operators from the
innermost to the outermost diagrams while for disjoined sub-diagrams
the expressions are naturally independent of the order of Taylor
operators, since then
\begin{equation}
I_{\Gamma}=I_{\Gamma \setminus \{ \gamma_1, \ldots, \gamma_n \}}
\prod_{k=1}^{n} I_{\gamma_{k}}.
\end{equation}
The forest formula then says that the integrand of the renormalized
diagram is given by the sum over all restricted forests:
\begin{equation}
R_{\Gamma}=\sum_{\mathfrak{F} \in \mathcal{F}_{R}(\Gamma)}
\Omega_{\mathfrak{F}}. 
\end{equation}
It is understood that the empty set stands for the diagram itself,
i.e., without any box around a sub-diagram.

The described BPHZ-scheme chooses the renormalization point for the
divergent diagrams at external momenta set to $0$. It is clear that by
another finite renormalization we can switch to any renormalization
scheme appropriate for the application considered. In our case of
$\phi^{4}$-theory we chose the on-shell scheme. We have to define the
coupling constant, the mass and the wave function normalization. This
can be formulated in terms of the proper self-energy and the proper
four-point vertex (the three-point vertex can be set identical to $0$
because of symmetry under field reflection $\phi \rightarrow -\phi$
without destroying the renormalizability of the theory, so that we do
not have to consider terms linear or cubic in the fields within the
Lagrangian):
\begin{equation}
\label{rencond}
\Gamma^{(4,\text{vac})}(s,t,u=0)=\frac{\lambda}{2}, \;
\Sigma^{(\text{vac})}(p^2=m^2)=0, \partial_{p^2} 
\Sigma^{(\text{vac})}(p^2=m^2)=0. 
\end{equation}
Here $s$, $t$, $u$ are the usual Mandelstam variables for two particle
scattering, $p$ is the external momentum of the self-energy and $m^2$
is the renormalized mass of the particles. The first condition defines
the coupling constant at vanishing momentum transfer for the
two-particle scattering to be given by $\lambda$, the second condition
chooses $m$ to be the physical mass of the particles, while the third
condition ensures that the residuum of the propagator at $p^2=m^2$ is
$1$ and thus the on-shell wave function is normalized to $1$ as it
should be.

\subsection{The finite temperature self-energy at the regularized level}

In this section we like to isolate those vacuum subparts inherent in
the pure temperature part of the self-energy which need to be
renormalized at physical space-time dimension $d=4$. For this purpose
we assume a regularization scheme, e.g., dimensional regularization
and extract those vacuum parts from the self-energy which diverge in
the limit $d\rightarrow 4$. For sake of clarification we mark all
equations with an asterix which diverge in the limit $d\rightarrow 4$
and which need special renormalization treatment. All other equations
are generally valid, even if all expressions are replaced by their
renormalized quantities.

In order to extract the divergent vacuum pieces we take the self-energy as
functional of the self-consistent propagator and expand it around the vacuum
value
\begin{equation}
\label{sig0}
\Sigma_{12}=\Sigma_{12}^{(\text{vac})} + 
\underbrace{\Sigma_{12}^{(0)} +
\Sigma_{12}^{(\text{r})}}_{\Sigma_{12}^{(\text{matter})}}
\end{equation}
Here $\Sigma_{12}^{(\text{vac})}$ is the vacuum ($T=0$) self-energy.
Its renormalization poses no particular problem and can be done
according to standard rules. At the examples discussed in our second
paper \cite{vHK2001-Ren-II} it is shown how to do this in practical
cases for the numerical solutions to the self-consistent equations of
motion. The second and third terms in (\ref{sig0}) contain the
in-matter or finite temperature components of the self-energy. Thereby
\begin{Equation*}
\label{sig0-def}
-\ii\Sigma_{12}^{(0)} = \funcint{\left ( \left .
      -\funcd{\ii\Sigma_{12}}{G_{1'2'}} \right|_{T=0}
    G_{1'2'}^{(\text{matter})} \right
  )}{1'2'}=\parbox{25mm}{\includegraphics{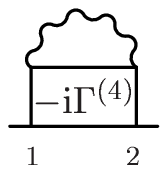}}
\end{Equation*}
contains the parts of $\Sigma$ linear in the matter (temperature) part
$G^{(\text{matter})}$ of the full propagator 
\begin{equation}
\label{GT}
\begin{array}{ccccc}
\ii G & =  & \ii G^{(\text{vac})}& +&  \ii G^{(\text{matter})} \\
\parbox{14mm}{\includegraphics{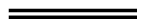}} & = &
 \parbox{14mm}{\raisebox{0.8mm}{\includegraphics{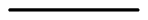}}}& + &
\parbox{15mm}{\includegraphics{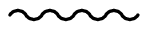}} 
\end{array} 
\end{equation}
At this level it is important to recognize that all loops involving
vertices from both sides of the real-time contour (cf.  Fig.
\ref{figure1}) are UV convergent due to the analytical properties of
$\{-+\}$ and $\{+-\}$-propagators (\ref{gmp}) to (\ref{a.15}). Such
loops contain at least one thermal weight factor $n(p_0)$ which decays
exponentially at large $|p_0|$, since the $\Theta$-function parts
completely drop out for large loop momenta.  Therefore all mixed
components like $G^{-+}$, or $\Sigma^{+-}$ have to be excluded from
the subtraction scheme. Thus, the expansion point, $G^{(\text{vac})}$,
in (\ref{GT}) is defined as the contour-diagonal part of the
propagator in the vacuum limit ($T\rightarrow 0$), i.e., with
vanishing \mbox{$G^{-+(\text{vac})}=G^{+-(\text{vac})}:=0$}. Likewise
all vacuum structures like $\Sigma^{(\text{vac})}$, and the four-point
functions $\Gamma^{(4,\text{vac})}$ and $\Lambda^{(\text{vac})}$ defined
below are ``diagonal'' in the real-time contour placement.

The remaining self-energy piece $\Sigma^{(\text{r})}$ in (\ref{sig0})
contains at least two $G^{(\text{matter})}$ lines which therefore are
never involved in any divergent loops due to the 2PI property of the
$\Phi$-functional\footnote{Any $G^{(\text{matter})}$-line in
  $\Sigma^{(\text{r})}$ is either involved in loops with further
  $G^{(\text{matter})}$-lines which are finite or it is attached to a
  pure vacuum piece. Due to the 2PI property of $\Phi$ this vacuum
  piece has more than four external legs also leading to finite loops
  for this $G^{(\text{matter})}$-line.}. Thus, there are no hidden
subdivergences in $\Sigma^{(\text{r})}$, and possible divergent vacuum
sub-structures can directly be renormalized using the BPHZ rules given
above.

On the other hand the diagrams of $\Sigma^{(0)}$ deserve special
attention, since there the single $G^{(\text{matter})}$-line is
involved in logarithmically divergent loops, if all vertices of
$\Gamma_{12,1'2'}^{(4)}$ are placed on the same side of the contour.
As mentioned the terms with mixed vertices are finite.

The divergences result from the fact that the
functional variation of $\Sigma$ with respect to $G$ at $T=0$ defines a
vacuum vertex function
\begin{equation}
\label{def-Ga}
-\ii \Gamma_{12,1'2'}^{(4)} =
-\left. \funcd{\Sigma_{12}}{G_{1'2'}} \right|_{T=0}
=\left.
-2\ii \frac{\delta^2 \Phi}{\delta G_{12}\delta G_{1'2'}}\right|_{T=0}
\end{equation}
with four external legs.

Its diagonal part (all vertices on one contour side) defines
$\Gamma_{12,1'2'}^{(4,\text{vac})}$ which is of divergence degree $0$.
Assuming $G^{(\text{matter})}$ of divergence degree $-4$ it is
involved in a logarithmically divergent loop. Thus, this part of
$\Sigma^{(0)}$, called $\Sigma^{(0,\text{div})}$, accounts for all
terms of divergence degree $0$ and consequently $\Sigma^{(\text{r})}$
is of divergence degree $-2$.

In order to trace all subdivergences hidden in $\Sigma^{(0)}$, the
vacuum structure inherent in $G^{(\text{matter})}$ has to be resolved.
For this purpose the diagonal parts of the full propagator
(i.e., $G^{--}$ and $G^{++}$) need further to be decomposed
\begin{equation}
\setlength{\unitlength}{1mm}
\label{Gr}
\ii G_{12}=\parbox{17\unitlength}{\centerline{%
$\ii G_{12}^{(\text{vac})}$}} +
\underbrace{\parbox{50\unitlength}{\centerline{%
$\displaystyle\ii\funcint{ G_{11'}^{(\text{vac})} 
  \Sigma_{1'2'}^{(0,\text{div})} G_{2'2}^{(\text{vac})}}{1'2'}$}}
+\parbox{17\unitlength}{\centerline{%
$\ii G_{12}^{\text{(r)}}$}}}_{\ii G^{(\text{matter})}} \hphantom{,}
\end{equation}
\begin{Equation*}
\setlength{\unitlength}{1mm}
  \hphantom{\ii G_{12}}=
  \parbox{17\unitlength}{\centerline{\includegraphics{Gvac.eps}}} +
  \parbox[b]{50\unitlength}{\centerline{\includegraphics{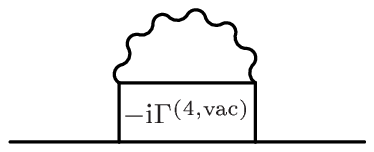}}} +
  \parbox{17\unitlength}{\includegraphics{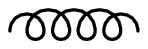}},
\end{Equation*}
where here and below all vertices $1,2,1'$ and $2'$ are placed on the
same side of the contour, this way defining the divergent piece of
$\Sigma^{(0)}$.  The remaining part $G^{(\text{r})}$ is of divergence
degree $-6$.

Since $\Gamma^{(4,\text{vac})}$ is a pure vacuum four-point function,
eqs. (\ref{sig0}) and (\ref{Gr}) also show $\Sigma^{(0,\text{div})}$
as a functional linear in $G^{(\text{matter})}$
\begin{Equation*}
\label{scsig0}
\Sigma_{12}^{(0,\text{div})}=\funcint{\Gamma_{12,1'2'}^{(4,\text{vac})}
  \left (\funcint{G_{1'1''}^{(\text{vac})}
      \Sigma_{1''2''}^{(0,\text{div})}
      G_{2''2'}^{(\text{vac})}}{{1''}{2''}} + G_{1'2'}^{(\text{r})}
  \right)}{1'2'}
\end{Equation*}
and thus also linear in $G_{1'2'}^{(\text{r})}$. A simple iteration
argument, starting with $\Sigma_{12}^{(0,\text{div})}=0$, shows that
Eq. (\ref{scsig0}) is solved by the ansatz
\begin{equation}
\label{defLa}
\setlength{\unitlength}{1.2mm} 
-\ii \Sigma_{12}^{(0,\text{div})}=
\funcint{\Lambda_{12,1'2'}^{(\text{vac})}
  G_{1'2'}^{(\text{r})}}{1'2'}=\parbox{15mm}{\centerline{%
\raisebox{3mm}{\includegraphics{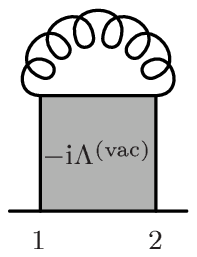}}}}
\setlength{\unitlength}{1.mm} 
\end{equation}
leading to the pure vacuum equation of motion
\begin{Equation*}
\label{Ladder}
\Lambda_{12,1'2'}^{(\text{vac})} = 
  \Gamma_{12,1'2'}^{(4,\text{vac})}+
\ii
\funcint{\Gamma_{12,34}^{(4,\text{vac})} 
  G_{35}^{(\text{vac})}G_{46}^{(\text{vac})}
  \Lambda_{56,1'2'}^{(\text{vac})}}{3456}
\end{Equation*}
for the vacuum four point function $\Lambda^{(\text{vac})}$. The
diagrammatic interpretation shows that this has the form of an
inhomogeneous Bethe-Salpeter ladder equation
\begin{Equation*}
\label{Ladder-diag}
\parbox{16.4mm}{\centerline{\includegraphics{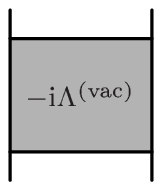}}} =
\parbox{16.4mm}{\centerline{\includegraphics{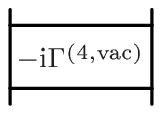}}} +
\parbox{16.4mm}{\centerline{\includegraphics{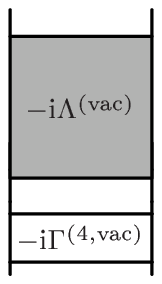}}} =
\parbox{16.4mm}{\centerline{\includegraphics{GammaFdiag.eps}}} +
\parbox{16.4mm}{\centerline{\includegraphics{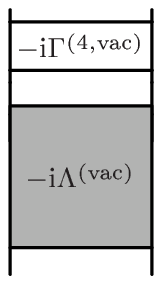}}}.
\end{Equation*}
From the construction it is clear that this is a very particular
BS-equation, namely the one which complies with the self-consistent
Dyson resummation scheme defined through (\ref{8c}). Thus these
ladders, which are of $s$-channel type (forward scattering), are
implicitly contained in the self-consistent vacuum self-energy. Since
Weinberg's power counting theorem shows that $\Lambda^{(\text{vac})}$
is of divergence degree $0$, as any four-point function, Eq.
(\ref{defLa}) implies that $\Sigma^{(0,\text{div})}$ is indeed also of
divergence degree $0$ as assumed above.  Then $G^{(\text{r})}$ is of
divergence degree $-6$ and the loops in (\ref{defLa}), which close
$\Lambda^{(\text{vac})}$ with a $G^{(\text{r})}$-line, contain no
further divergences. Thus, once $\Lambda^{(\text{vac})}$ is
renormalized, the self-energy is renormalized too, within a
temperature independent subtraction scheme.

Since $\Phi$ is 2PI, the variational relation (\ref{def-Ga}) defines
the BS-kernel $\Gamma_{12,1'2'}^{(4,\text{vac})}$ as a proper skeleton
diagram, i.e., it contains no self-energy insertions and cutting the
diagram such that the pairs of space time points $(12)$ and $(1'2')$
are separated, cuts more than two lines.  Thus the BS-kernel has the
appropriate irreducibility properties for the resummation to the
complete four-point function $\Lambda^{(\text{vac})}$, again showing
the virtue of the $\Phi$-functional formalism to avoid double
counting.

From (\ref{defLa}) it is obvious that switching to the momentum space
representation both $\Gamma^{(4,\text{vac})}$ and
$\Lambda^{(\text{vac})}$ are not needed in their full momentum
dependence but rather only as a function of the two momenta given by
the Fourier transformation with respect to the space-time point pairs
$(12)$ and $(1'2')$. Due to (\ref{def-Ga}) and through (\ref{Ladder})
both, $\Gamma^{(4,\text{vac})}$ and $\Lambda^{(\text{vac})}$, obey the
symmetry relations
\begin{equation}
\label{symm-ga}
\begin{array}{rclcrcl}
\Gamma_{12,1'2'}^{(4,\text{vac})}&=&\Gamma_{1'2',12}^{(4,\text{vac})}&
\quad\mbox{or}\quad&\Gamma^{(4,\text{vac})}(p,q) &=&
\Gamma^{(4,\text{vac})}(q,p), \\
\Lambda_{12,1'2'}^{(\text{vac})}&=&\Lambda_{1'2',12}^{(\text{vac})}&
\quad\mbox{or}\quad&\Lambda^{(\text{vac})}(p,q) &=&
\Lambda^{(\text{vac})}(q,p).
\end{array}
\end{equation}

\subsection{Renormalization of the vacuum Bethe-Salpeter equation}

In energy-momentum representation the regularized BS-equations
(\ref{Ladder}) and its equivalent ``adjoint'' version become
\begin{Eqnarray*}
\label{mom-ladder}
\Lambda^{(\text{vac})}(p,q)&=&\Gamma^{(4,\text{vac})}(p,q)+\ii \feynint{l}
\Gamma^{(4,\text{vac})}(p,l) [G^{(\text{vac})}(l)]^2
\Lambda^{(\text{vac})}(l,q)\\  
&=&\Gamma^{(4,\text{vac})}(p,q)+\ii \feynint{l}
\Lambda^{(\text{vac})}(p,l) [G^{(\text{vac})}(l)]^2
\Gamma^{(4,\text{vac})}(l,q).
\end{Eqnarray*}
The renormalization of the BS-equations (\ref{Ladder}) is not straight
forward. First the BS-kernel $\Gamma^{(4,\text{vac})}(p,q)$ has to be
renormalized following the BPHZ-rules outlined in sect.  \ref{sect3}.
Representing $\Lambda(p,q)$ as the sum of ladder diagrams, this
BS-kernel forms the rungs in each ladder. The complication arises from
overlapping sub-divergences: in the general case each internal rung is
part of various diverging sub-diagrams through the loops involving two
or more rungs. However, two observations help to settle this
renormalization issue:

\vspace{0.5mm}
\begin{minipage}{0.7\textwidth}
\begin{itemize}
\item[(i)] The rungs given by the BS-kernel are 2PI with respect to
  cuts separating the top from the bottom extremities of the rung.
  This implies that there are no divergent sub-diagrams which cut into
  the inner structure of any rung, since this would involve a cut of
  more than two lines and the resulting sub-diagram would have more
  than four external lines.  Thus divergent sub-diagrams always have
  complete rungs as sub-diagrams, as shown to the right.
\item[(ii)] If one takes the difference of two $\Lambda$ functions
  which differ only in one of the momenta, e.g., $\Lambda(p,q) -
  \Lambda(p',q)$, all those counter terms cancel out which contain
  boxes which cut the outer $p$ or $p'$ lines, respectively, since for
  these counter terms the argument $p$ or $p'$ are replaced by zero.
  Thus, the only boxes which are left are those which exclude the
  outermost rung attached to the $p$-lines.  These boxes, however,
  just define the renormalized result of the sub-diagram complementary
  to this outermost rung, which is again a ladder diagram.
\end{itemize}
\end{minipage}
\begin{minipage}{0.25 \textwidth}
\centering{\includegraphics{LadderBox.eps}}\\[3mm] 
(valid counter term)
\end{minipage}
\vspace{0.5mm}

This permits to establish recursive relations for the renormalized
expressions of the two possible differences
\begin{align}
\label{diff1-ren}
\Lambda^{(\text{ren})}(p,q)-\Lambda^{(\text{ren})}(p',q)  &
\Gamma^{(4,\text{vac})}(p,q) -\Gamma^{(4,\text{vac})}(p',q)  \\
\nonumber
& + \ii \fint{l}
[\Gamma^{(4,\text{vac})}(p,l) -\Gamma^{(4,\text{vac})}(p',l)]
[G^{(\text{vac})}(l)]^2 \Lambda^{(\text{ren})}(l,q),\\
\label{diff2-ren}
\Lambda^{(\text{ren})}(p',q)-\Lambda^{(\text{ren})}(p',q')  &
\Gamma^{(4,\text{vac})}(p',q) -\Gamma^{(4,\text{vac})}(p',q')  \\
\nonumber
& + \ii \fint{l} \Lambda^{(\text{ren})}(p',l)[G^{(\text{vac})}(l)]^2
[\Gamma^{(4,\text{vac})}(l,q) -\Gamma^{(4,\text{vac})}(l,q')],
\end{align}
where now $\Gamma^{(4,\text{vac})}$ stands for the renormalized
Bethe-Salpeter kernel. Since the renormalized function
$\Gamma^{(4,\text{vac})}(p,l)-\Gamma^{(4,\text{vac})}(0,l)$ is of
divergence degree less than zero the integrals are finite. This set of
renormalized equations can be used to construct the renormalized
$\Lambda$-function using
\begin{equation}
\label{diff2}
\Lambda^{(\text{ren})}(0,0) := \pm \frac{\lambda}{2}
\end{equation}
on the two real-time branches $\mathcal{C}_{\mp}$ due to our
renormalization condition (\ref{rencond}). In a kind of sweep-up
sweep-down scheme first the ``half sided''
$\Lambda^{(\text{ren})}(0,q)$ function can be constructed by solving
(\ref{diff2-ren}) for $p'=0$.  Using this half sided function as the
input for (\ref{diff1-ren}) the full momentum dependence of
$\Lambda^{(\text{ren})}(p,q)$ can be obtained. This scheme fully
complies with the BPHZ renormalization prescription. It has the
remarkable feature that, although it is explicitly asymmetric in $p$
and $q$, it constructs a completely symmetric renormalized four-point
function which can be combined to the complete result
\begin{equation}
\label{La-ren}
\begin{split}
  \Lambda^{(\text{ren})}(p,q)= & \Gamma^{(4,\text{vac})}(p,q) \\ & +
  \ii \fint{l} [\Gamma^{(4,\text{vac})}(p,l)
  -\Gamma^{(4,\text{vac})}(0,l)] [G^{(\text{vac})}(l)]^2
  \Lambda^{(\text{vac})}(l,q) 
  \\
  &+ \ii \fint{l} \Lambda^{(\text{ren})}(0,l) [G^{(\text{vac})}(l)]^2
  [\Gamma^{(4,\text{vac})}(l,q) -\Gamma^{(4,\text{vac})}(l,0)].
\end{split}
\end{equation}
For numerical applications it is important to realize that only the
half sided $\Lambda^{(\text{ren})}(0,l)$, and not the full momentum
dependence $\Lambda^{(\text{ren})}(p,q)$, is explicitly needed. Since
the half side $\Lambda$ has essentially two-point function properties
it can be numerically constructed using similar techniques as for
self-energies.  

Indeed one can express the complete renormalized self-energy part
  linear in $G^{(\text{matter})}$ in the form
\begin{equation}
\setlength{\unitlength}{1.2mm} 
\label{sigren-simple}
\begin{split}
  \Sigma^{(0)}(p) & =\Sigma^{(0)}(p)-\Sigma^{(0)}(0)+\Sigma^{(0)}(0) \\
  & = \fint{l} [\Gamma^{(4)}(p,l)-\Gamma^{(4,\text{vac})}(0,l)]
  G^{(\text{matter})}(l)  \\
  & \quad + \fint{l} \Lambda^{(\text{ren})}(0,l) G^{(\text{r})}(l)
  \\
  & = \underbrace{
    \parbox{25mm}{\raisebox{6mm}{\includegraphics{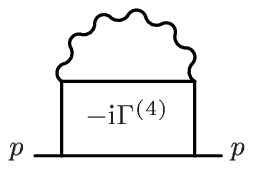}}} -
    \parbox{25mm}{\raisebox{6mm}{\includegraphics{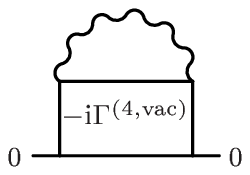}}}}_{
    \text{finite}} +
  \parbox{25mm}{\raisebox{6mm}{\includegraphics{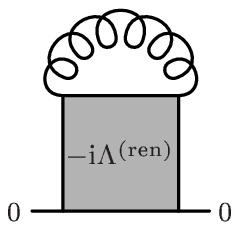}}}
\end{split}
\setlength{\unitlength}{1mm} 
\end{equation}
with $G^{(\text{matter})}$ and $G^{(\text{r})}$ from (\ref{GT}) and
(\ref{Gr}). Here $\Gamma^{(4)}$ is the full contour valued kernel
(\ref{def-Ga}) including mixed contour vertices, while
$\Gamma^{(4,\text{vac})}$ and $\Lambda^{(\text{ren})}$ are diagonal in
the contour vertices.  Due to the 2PI-properties of $\Gamma^{(4)}$ the
difference $\Gamma^{(4)}(l,p)-\Gamma^{(4,\text{vac})}(l,0)$ is of
divergence degree less than $0$.  Therefore the first integral
represented by the difference of the first two diagrams is finite,
since $G^{(\text{matter})}$ is of divergence degree $-4$. This
difference represents the most naive subtraction, which by itself,
however, would be false, since it contains temperature dependent
counter terms. The heart of the above derivation is that these false
$T$-dependent counter terms are precisely compensated by the last
term. The fact that the counter term structures never mix the two
real-time contour branches also lifts the problem of pinch
singularities which otherwise could arise due to the vanishing
external momentum.

This completes the proof that the self-energies can be renormalized
with $T$-independent counter terms.

\subsection{Renormalization of the real-time $\Gamma$-functional}
\label{sect-Re-Ga}

In this section we derive the renormalized real-time generating
functional for the in-matter equations of motion. Thus, we restrict
the contour integrations to the real-time contour $\CR$ and the
corresponding traces to the real-time traces $\TrR$. For functions in
momentum representation the corresponding contour matrix algebra in
the $\{-+\}$ notation (cf. Appendix \ref{appa}) is implied.

For the renormalization procedure we use the ansatz (\ref{Gr}) for the
full propagator $G$ together with the form (\ref{defLa}) for the
logarithmically divergent part of the self-energy, where the
renormalized four-point function $\Lambda$ resolves the subdivergences
hidden in both, the propagator and the self-energy.  For this purpose
we decompose the generating functional $\Gamma$ in its vacuum part,
which is solely a functional of $G^{(\text{vac})}$, and the in-matter
part
\begin{equation}\label{Gamma-vac+matter}
\Gamma=\Gamma^{(\text{vac})}[G^{(\text{vac})}]+
\Gamma^{(\text{matter})}[G^{(\text{vac})},G^{(\text{matter})}]
\end{equation}
Thereby it is implied that the vacuum problem is already solved through its 
equation of motion resulting from the functional variation of 
$\Gamma^{(\text{vac})}$. Given $G^{(\text{vac})}$ the equations of motion in 
matter result from the functional variation of $\Gamma^{(\text{matter})}$ with 
respect to $G^{(\text{matter})}$.

Compared to the two-point self-energy the $\Gamma$- and
$\Phi$-functionals have no external points and essentially result from
the diagrams of the self-energies by closing the extremities.
Therefore one has to explicitly expand the corresponding expressions
up to second order in $G^{(\text{matter})}$ before one comes to the situation
where the remaining pieces are void of hidden subdivergences. Thus we write
\begin{equation}
\begin{split}
  \Phi_{\R} & = \Phi_{\R}^{(\text{vac})} + \TrR
  \funcd{\Phi^{(\text{vac})}}{G^{(\text{vac})}} G^{(\text{matter})} 
  + \frac{1}{2!}
  G^{(\text{matter})} \frac{\delta^2 \Phi^{(\text{vac})}}{\delta
    {G^{(\text{vac})}}^2}
  G^{(\text{matter})} + \Phi_{\R}^{(\text{r})} \\
  &= \Phi_{\R}^{(\text{vac})} - \frac{\ii}{2} \TrR G^{(\text{matter})}
  \Sigma^{(\text{vac})} + \frac{1}{4} G^{(\text{matter})}
  \Gamma^{(4)} 
  G^{(\text{matter})} + \Phi_{\R}^{(\text{r})}.
\end{split}
\end{equation}
Here we have used (\ref{8c}) and (\ref{def-Ga}) for the vacuum parts
defined through the variation of $\Phi$ with respect to $G$. At the
same time we introduced the real-time trace in momentum space
\begin{equation}\label{R-trace}
\TrR A \cdots B=\feynint{l} A(l) \cdots B(l)
\end{equation}
and the functional tensor
contraction for four-point functions with propagators:
\begin{equation}
G_{1} \Gamma^{(4)} G_{2} = \feynint{l_1} \feynint{l_2} G_{1}(l_1)
\Gamma^{(4)}(l_1,l_2) G_{2}(l_2).
\end{equation}
In all expressions the functions are contour matrix functions which
imply the corresponding contour matrix algebra and the contour
trace, cf. Eqs. (\ref{CR-Trace}) and (\ref{CRp-Trace}).

Applying the arguments given for $\Sigma^{(\text{r})}$ also for
$\Phi^{(\text{r})}$ no $G^{(\text{matter})}$-line is involved in
divergent loops such that after renormalization of possible vacuum
sub-divergences the entire diagram is finite. Thus only terms with at
most two $G^{(\text{matter})}$-lines need further care.

Since $\Phi$ by itself is not an observable we directly step to the
definition of the $\Gamma$-functional which relates to the
thermodynamic potential in the equilibrium case. Exploiting the
stationarity condition of $\Gamma$ at the vacuum level, i.e., using
the vacuum equations of motion for $G^{(\text{vac})}$, all terms
linear in $G^{(\text{matter})}$ drop out and we find for the
functional with all proper vacuum sub-divergences subtracted
\begin{Equation*}
\begin{split}
\label{Gammabar}
\bar{\Gamma}_{\R}^{(\text{matter})}[G^{(\text{matter})}]
=&\frac{\ii}{2} \TrR \left (
  G^{(\text{matter})} \Sigma^{(\text{matter})} - \sum_{k=2}^{\infty}
  \frac{(G^{(\text{vac})} \Sigma^{(\text{matter})})^{k}}{k} \right)\\ 
&-\frac{\ii}{4} \TrR G^{(\text{matter})} \Sigma^{(0)} +
\bar{\Phi}_{\R}^{(\text{r})}.
\end{split}
\end{Equation*}
Renormalized to zero at the vacuum level this expression is a
functional of the in-matter part of the propagator. The remaining
divergent parts of $\bar{\Gamma}^{(\text{matter})}$ arise from terms
quadratic in 
$G^{(\text{matter})}$, 
i.e.,
\begin{Equation*}\label{Gammabar-div}
\bar{\Gamma}_{\R}^{(\text{matter,div})} = \frac{\ii}{4} \TrR G^{(0)}
\Sigma^{(0)} + \frac{1}{4} G^{(0)} \Gamma^{(4,\text{vac})} G^{(0)}
+\frac{1}{2} G^{(0)} \Gamma^{(4,\text{vac})} G^{(\text{r})}
\end{Equation*}
with
\begin{equation}
\label{G0}
G^{(0)}=G^{(\text{vac})}\Sigma^{(0,\text{div})}G^{(\text{vac})},
\end{equation}
where again in both relations above all expressions are contour
diagonal. Both, $\Sigma^{(0,\text{div})}$ and $G^{(0)}$ are
linear in $G^{(\text{matter})}$.  Using the equations of motion for
$\Sigma^{(0,\text{div})}$ and $\Lambda^{(\text{vac})}$ one
arrives at an expression for the divergent part of
$\bar{\Gamma}^{(\text{matter})}$ which only contains quantities which
were already renormalized in the previous subsection
\begin{equation}
\Gamma_{\R}^{\text{(matter,div,ren)}} = \frac{1}{4} \left (G^{(\text{r})}
  \Lambda^{(\text{ren})} G^{(\text{r})} - G^{(\text{r})}
  \Gamma^{(4,\text{vac})} G^{(\text{r})} \right).
\end{equation}
Substituting this for the divergent part we obtain after some
algebraic simplifications
\begin{equation}
\label{Gamma-ren}
\begin{split}
  \Gamma_{\R}^{\text{(matter,ren)}}[G^{(\text{matter})}] = &
  \frac{\ii}{2} \overline{\TrR} \Bigg ( G^{(\text{r})}
  \Sigma^{(\text{matter})}-\frac{1}{2} G^{(\text{vac})}
  \Sigma^{\text{(r)}} G^{(\text{vac})} \Sigma^{(\text{r})}  \\ &
  \quad - \sum_{k=3}^{\infty} \frac{(G^{(\text{vac})}
    \Sigma^{(\text{matter})})^{k}}{k} \Bigg ) + \frac{1}{4}
  G^{(\text{r})} \Lambda^{(\text{vac})} G^{(\text{r})} +
  \Phi_{\R}^{(\text{r})}\\
  &+ \text{mixed contour $\TrR$-terms from (\ref{Gammabar})},
\end{split}
\end{equation}
where $\overline{\TrR}$ includes only the contour diagonal parts.
This expression, which now can be considered as a functional of
$G^{(\text{matter})}$, or through Eqs. (\ref{Gr}) and (\ref{defLa}) of
$G^{(\text{r})}$, is void of any hidden subdivergences, since all
matter or $T$-dependent parts of the propagator like
$G^{(\text{matter})}$ or $G^{(\text{r})}$ are involved in convergent
loops.

Now it remains to be proven that this renormalization procedure for
the $\Gamma$-functional is consistent with the renormalization of the
self-energy given in the previous section. In other words: We like to
show that the vanishing functional variation of
$\Gamma^{\text{(ren)}}[G^{(\text{matter})}]$ complies with the Dyson
equation of motion and the renormalized self-energy. From the
BPHZ-formalism we expect this to hold true, because it ensures that
sub-divergences can be renormalized first and then the remaining
divergences which come into the game by closing the diagrams: The
renormalized result is independent of the order of counter term
subtractions.

It is sufficient to show this for the contour diagonal parts of
(\ref{Gamma-ren}). Writing the functional variation of
$\Gamma^{\text{(ren)}}$ as
\begin{equation}\label{varGamma-ren}
\delta \Gamma_{\R}^{\text{(matter,ren)}}=
\funcd{\Gamma_{\R}^{\text{(matter,ren)}}}{\Sigma^{(\text{r})}}
\delta\Sigma^{(\text{r})}+
\funcd{\Gamma_{\R}^{\text{(matter,ren)}}}{G^{(\text{r})}} \delta
G^{(\text{r})}, 
\end{equation}
where $\Sigma^{(\text{r})}$ is supposed to be a functional of
$G^{(\text{r})}$. Both terms in (\ref{varGamma-ren}) independently
vanish. The first term drops by virtue of the Dyson equation which
together with (\ref{Gr}) ensures
\begin{equation}
\delta\left({\sum_{k=3}^{\infty}
  \frac{(G^{(\text{vac})} \Sigma^{(\text{matter})})^{k}}{k}}\right)=
\left(G^{(\text{r})}-G^{(\text{vac})}\Sigma^{(\text{r})}G^{(\text{vac})}
\right)\delta \Sigma^{(\text{matter})}.
\end{equation}
The second term shrinks to
\begin{equation}
\delta \Gamma_{\R}^{\text{(matter,ren)}}=\frac{\ii}{2}\TrR
\underbrace{
\left(\Sigma^{(\text{r})}\delta G^{(\text{r})}+
G^{(\text{vac})}\Sigma^{(\text{r})}G^{(\text{vac})}\delta\Sigma^{(0)}
\right)}_{\Sigma^{(\text{r})}\delta G^{(\text{matter})}}
+\delta\Phi_{\R}^{(\text{r})} \stackrel{!}{=}0,
\end{equation}
which indeed implies
\begin{equation}
\Sigma^{(\text{r})}=
2 \ii \funcd{\Phi_{\R}^{(\text{r})}}{G^{(\text{matter})}},
\end{equation}
compatible with the definition of $\Sigma^{(\text{r})}$. It is
important to note that through the functional variation
(\ref{varGamma-ren}) only convergent loops are opened, such that none
of the counter terms is affected by this variation. This explicitly
demonstrates the consistency of the BPHZ-renormalization scheme for
the self-consistent approximations: The operations of variation with
respect to $G$ and renormalization are commutative, i.e., one can
construct the renormalized self-energy in two equivalent ways: The
first uses the \emph{un-renormalized} $\Phi$-functional and defines
the renormalized self-energy by applying the BPHZ-renormalization
theorem to its diagrams which are defined by opening any line of the
un-renormalized $\Phi$-functional. In this way we have defined the
renormalized self-energy in the previous section.  Subsequently we
renormalized the $\Gamma$-functional by substituting the thereby
defined renormalized functions for the divergent vacuum-sub-diagrams
with $2$ and $4$ external legs. After subtracting the pure vacuum
contribution this leads to the finite renormalized functional
(\ref{Gamma-ren}).

As we have seen now, the second way to define the renormalized
self-energy is to renormalize the $\Gamma$-functional first. Then the
variation with respect to the \emph{renormalized} propagator leads to
the \emph{renormalized} equations of motion and thus directly to the
\emph{renormalized} self-energy which shows the consistency of the
local vacuum counter terms including all combinatorial factors for
both, the generating functional and the equation of motion.

We like to clarify that the equations derived in this section for the
in-matter parts of the propagator are all valid also in the general
non-equilibrium case of quantum field theory provided the density
operator at time $t_0$ amends a Wick decomposition. This is valid for
statistical operators at initial time of the form \cite{dan84}
\begin{equation}
\op{R}=\frac{1}{Z}\exp(-\sum_{k} \alpha_{k} \op{A}_{k}) \text{ with }
Z=\Tr \exp(-\sum_{k} \alpha_{k} \op{A}_{k}) , 
\end{equation}
where the $\op{A}_{k}$ are one-particle operators. Our arguments for
the renormalizability with $\alpha_{k}$ independent (i.e., state
independent) local counter terms should hold, since the statistical
operator is normalized $\Tr \op{R}=1$. Thus, any in-matter part of the
propagator leads to similar reductions of the degree of divergence as
for the Bose-Einstein distribution functions used here. This ensures
that the power counting arguments for the non-vacuum parts are still
valid.  Note in particular that the real-time functional $\Gamma$ has
only functional meaning, namely as a tool to derive the equations of
motion, since its value at the physical solution vanishes.

\subsection{Renormalization of the thermodynamical potential}

Using the thermodynamic part, i.e., the vertical branch, of the contour
given in Fig. \ref{figure1} the $\Gamma$-functional provides a finite
value which indeed relates to the thermodynamic potential $\Omega$.
For the evaluation one uses the relationship between the Matsubara
functions and the real-time functions given in Appendix \ref{appa},
Eq. (\ref{a.14}) for the thermodynamic trace $\TrT$ which takes due
account of the thermodynamic weights in the partition sum
\begin{equation}
\label{T-trace}
\begin{split}
\TrT\{h(p)\}&= \beta V\feynint{p}\left(h^{-+}+h^{+-}\right)\\
&= -2\ii\beta V\feynint{p} \sign(p_0)\left(n(p_0)+\frac12\right)
\im h_R(p),
\end{split}
\end{equation}
where $h^{-+}$ and $h^{+-}$ are the Wightman functions  and $h_R$ 
is the retarded function of $h$. Furthermore
\begin{equation}
\label{BE}
n(p_0)=\frac{1}{\exp(\beta|p_0|)-1}
\end{equation}
is the thermal Bose-Einstein factor resulting from the summation over
Matsubara frequencies expressed in terms of complex contour integrals
cf. (\ref{a.14}). For this thermal contour $\CT$ closed diagrams as
those of $\Phi$ and $\Gamma$ also attain a finite value. In this case
the rule is, first to omit one of the momentum integrations, which in
this way defines a two-point function. Its renormalized retarded value
can be calculated according to the above used real-time contour rules.
Subsequently one applies (\ref{T-trace}) for the final
integration. We show now that after renormalization of the two-point
function we need only to subtract the overall vacuum divergence
inherent in this final integral.

To obtain this result we have to go back to the regularized expression
for the \emph{un-renormalized} effective potential (\ref{7}) which
relates to the thermodynamical potential via
\begin{equation}\label{Omega-reg}
\Omega^{(\text{reg})}(T)=-T\Gamma^{(\text{reg})}_{\text{Th}}
\end{equation}
where 
\begin{equation}
\begin{split}
\label{GammaT}
\Gamma^{(\text{reg})}_{\text{Th}}(T)=& 
 \Gamma^{(\text{vac,reg})}_{\text{Th}}(T)
+\Gamma^{(\text{matter,reg})}_{\text{Th}}(T) \quad\quad\text{with}\\
\Gamma^{(\text{vac,reg})}_{\text{Th}}(T)=&
\TrT\left[  \frac{\ii}{2} \ln\left(-\frac{M^2}{ G^{\text{vac}}}\right) 
+ \frac{\ii}{2} \Sigma^{\text{vac}}
  G^{\text{vac}}\right]  +\Phi_{\text{Th}}^{\text{vac}}(T),\\ 
 \Gamma^{(\text{matter,reg})}_{\text{Th}}(T)=&
  \frac{\ii}{2} \TrT \left [ G^{(\text{r})}
  \Sigma^{(\text{matter})}-\frac{1}{2} G^{(\text{vac})} \Sigma^{\text{(r)}}
  G^{(\text{vac})} \Sigma^{(\text{r})} - \sum_{k=3}^{\infty}
  \frac{(G^{(\text{vac})} \Sigma^{(\text{matter})})^{k}}{k} \right] \\
   & + \frac{1}{4}
  G^{(\text{r})} \Lambda^{(\text{vac})} G^{(\text{r})} 
  +\Phi_{\text{Th}}^{(\text{r})}(T).
\end{split}
\end{equation}
Here the subscript $\text{Th}$ specifies the quantities resulting from
the thermal trace. The matter part results from form
(\ref{Gamma-ren}).  Subsequently one replaces all quantities by their
renormalized ones (denoted by a bar across the functions) and cancels
the overall divergence by subtracting the $T\rightarrow +0$ value
\begin{equation}\label{Omega-ren}
\begin{split}
\Omega^{(\text{ren})}(T)
&=-T\left(
\bar{\Gamma}^{(\text{reg})}_{\text{Th}}(T)
-\bar{\Gamma}^{(\text{reg})}_{\text{Th}}(+0)\right)\\
&=-T\left(\bar{\Gamma}^{(\text{vac,reg})}_{\text{Th}}(T)
-\bar{\Gamma}^{(\text{vac,reg})}_{\text{Th}}(+0)
+\bar{\Gamma}^{(\text{matter,reg})}_{\text{Th}}(T)\right).
\end{split}
\end{equation}

This procedure is legitimate as long as the new loops due to the final
thermodynamical trace (\ref{T-trace}) do not induce new
subdivergences, but only overall divergences. It is obvious that for
the matter part the final trace loop involving the factor $n+1/2$ is
completely convergent, since all loops are regular once $\Lambda$ is
renormalized. For the two vacuum terms the component proportional to
the factor $1/2$ in the thermal trace (\ref{T-trace}) cancel out, such
that all terms are proportional to $n(p_0)$ which cuts off the $p_0$
integration, while the momentum integrations are also limited due to
the vacuum thresholds: the imaginary parts of vacuum functions are
zero for $p^{2}<m^{2}$, where $m$ is the mass of the stable vacuum
particle. Thus also these final loop integrals are finite, defining a
finite thermodynamical potential.

The vacuum part essentially determines the kinetic energy part of
$\Omega$ as can be seen from the most simple example of an ideal gas.
Here of course all self-energies and $\Phi$ are vanishing, the
retarded propagator at finite temperature is $D_R(p)=[p^2-m^2+\ii \eta
\sigma(p_0)]^{-1}$ and the renormalization is done by subtracting the
pure vacuum part.  Thus, the free thermodynamical potential becomes
\begin{equation}
\label{Gammaidgasone}
\Omega^{(\text{id. gas})}=- V \fint{l} n(l_0) \pi \Theta(l^2-m^2),
\end{equation}   
which can be brought to a more familiar form by an integration by
parts
\begin{equation}
\label{Gammaidgastwo}
\Omega^{(\text{id. gas})}=- p V =  V \int \frac{\d^{3} \vec{l}}{(2
  \pi)^3} \ln \left [1-\exp \left ( -\beta \sqrt{\vec{l}^2+m^2}
  \right) \right]. 
\end{equation}

\section{Conclusions and outlook}
\label{sect6}

For the example of $\phi^4$-theory we have shown that self-consistent
Dyson resummations based on a $\Phi$-derivable scheme can be
renormalized with local counter terms defined on the self-consistently
determined vacuum level. This result was obtained with help of
Weinberg's power counting theorem and using the BPHZ-renormalization
scheme with the usual modifications for finite temperature diagram
rules, which can be summarized in the simple rule that the
``contraction boxes'' defining the counter terms have to exclude
sub-diagrams which contain any temperature line.

The hidden subdivergence structure of the self-consistent scheme has
been resolved. This leads to a Bethe-Salpeter equation for the vacuum
four-point function compatible with the chosen $\Phi$-approximation,
which we have renormalized. The method is free of pinch singularities.
Closed equations could be formulated which resum the non-perturbative
structure of both, the equations of motion, i.e., the self-energies
and also the non-perturbative counter-term structure. The complexity
of these equations is comparable to standard Dyson resummation schemes
and therefore in principle does not imply new techniques. First
numerical applications, which include the construction of the
BS-kernel, the solution of the half-sided four-point function and thus
the renormalized self-energies up to the self-consistent sunset
self-energy, are presented in a second paper \cite{vHK2001-Ren-II}.
The renormalization of the generating functional $\Gamma$, c.f. sect.
\ref{sect-Re-Ga}, shows that the derivation and thus the renormalized
in-matter equations of motion equally apply to the general
non-equilibrium case.

This also proves that there is no arbitrariness in studying the
in-medium modifications of model parameters like the mass and the
coupling constants within this class of approximation schemes: It is
sufficient to adjust them in the vacuum, for instance by fitting them
to scattering data, in order to predict without ambiguity how they
change in the dense and hot medium: \emph{The in-medium modifications
  are ruled completely by the model and its vacuum parameters alone,
  no further assumptions need to be made.}

Although demonstrated for the $\phi^4$-theory, the method is in
principle general, since the derivation only relies on the analytic
and asymptotic form of the propagators. In particular the
renormalization of hidden overlapping divergences in the
logarithmically divergent Bethe-Salpeter equations is general.  Still,
there is a number of restrictions of the self consistent Dyson
resummation within the $\Phi$-derivable scheme, which concerns global
and local symmetries and the corresponding conservation laws and
Ward-Takahashi identities.

The $\Phi$-functional formalism only ensures the conservation laws for
the expectation values of charges associated with the symmetry by
Noether's theorem. However, in general the Ward-Takahashi identities
are violated for the self-energy and higher vertex functions.
Heuristically the problem can be traced back to the violation of
crossing symmetry by the self-consistent scheme: Our derivation shows
that the self-consistent solution of the self-energy involves
Bethe-Salpeter ladder resummations of the four-point function, but
only in the $s$-chanel. The crossing symmetric $t$- and $u$- channel
contributions to the four-point function are not included.

The symmetry properties of the $\Gamma[\varphi,G]$-functional were
already investigated by us with help of the here applied path-integral
method \cite{vH-Ph-D00,vHK2001-Ren-III}. We show that it is always
possible to define a \emph{non-perturbative} approximation to the
effective action $\Gamma_{\text{eff}}[\varphi]$ which respects
linearly realized symmetries of the classical action provided the
symmetry is not anomalously broken. The self-energy and higher vertex
functions defined from this improved approximation action formalism
then fulfill the Ward-Takahashi identities of the underlying symmetry.
As a result the effective action $\Gamma_{\text{eff}}$ enforces that
additional $t$- and $u$-channel Bethe-Salpeter resummations are needed
to restore the crossing symmetry together with the Ward-Takahashi
identities for the self-energy and the vertex functions. However,
these vertex functions are not self-consistently calculated and thus
some problems remain also within this approximation: For instance in
the case of the linear sigma-model the O(N)-symmetry is restored for
the vertex-functions and the Goldstone-modes become massless. Yet, the
phase transition from the Nambu-Goldstone phase at low temperatures to
the Wigner-Weyl phase at high temperatures results to be of
$1^{\text{st}}$ order rather than $2^{\text{nd}}$ order
\cite{baymgrin}.
  
In the case of a local gauge symmetry the problems become even more
intricate: Self-consistent schemes beyond the classical field level
for the gauge fields generally violate local gauge symmetries for the
same reasons as for global symmetries.  However, this immediately
causes the excitation of spurious modes of the gauge fields which
leads to violation of the unitarity of the S-matrix, the positive
definiteness of the statistical operator and the causality structure
of Green's functions. Nevertheless a gauge invariant effective action
$\Gamma_{\text{eff}}$ within a background-field approach can be
formulated which provides gauge covariant polarization functions
\cite{vHK2001}.

From a practical point of view the problem remains to calculate the
self-consistent propagators needed for the symmetry-restoring
Bethe-Salpeter resummation, which presently can only be solved in
simple cases (RPA bubble resummation). In \cite{vHK2001} we have
presented a workaround in terms of a suitably chosen projection
method onto the physical (transverse) degrees of freedom of the
gauge-field polarization tensor. This procedure, of course, does not
lead to a full restoration of local gauge theory but to causal Green's
functions and current conservation within the self-energies of
matter-fields. Alternative methods are restricted to the approximate
solution of the self-consistent equations of motion, e.g., in the
sense of a Hard thermal loop approximation \cite{pesh00} or to a
systematic expansion in terms of the coupling constant or $\hbar$
\cite{ianc00}.

The proof of the renormalizability of $\Phi$-derivable approximations
opens a broad range of perspectives for effective field theory model
applications describing the non-perturbative in-medium properties of
particles in dense or finite-temperature matter with model parameters
fixed at the vacuum level. Further applications point towards the
appropriate renormalization of non-equilibrium transport equations
\cite{KIV01}, where in particular the drift terms, which determine the
equation of state, involve the real part of the self-energies which
generally need renormalization.

\section*{Acknowledgments}

We are grateful to G. E. Brown, P. Danielewicz, B. Friman, Yu. Ivanov,
E. E.  Kolomeitsev, M. Lutz, M. A. Nowak and D. Voskresensky for
fruitful discussions and suggestions at various stages of this work.

\appendix

\section{Analytical properties of Green's functions}
\label{appa}

In this appendix we summarize briefly the analytic properties of
Green's functions of neutral bosons needed in the main part of the
paper. This is most easily done by switching to the operator formalism
in the Heisenberg picture. By definition we have for a hermitian
scalar field operator
\begin{equation}
\label{a.1}
\ii G(x) = \frac{1}{Z} \Tr \exp(-\beta \op{H})  \op{\phi}(x)
\op{\phi}(0):=\erw{\mathcal{T}_{\mathcal{C}} \op{\phi}(x)
\op{\phi}(0)}_{\beta}. 
\end{equation}
For $x^0$ on the vertical part, i.e., $x^0=-\ii \tau$ with $0 \leq \tau
\leq \beta$ we obtain the Matsubara Green's function
\begin{equation}
\label{a.2}
G_M(\tau,\vec{x})=G^{+-}(-\ii \tau,\vec{x}),
\end{equation}
where one has to understand the analytic continuation of the real-time
Wightman function $G^{+-}$ on the right hand side. It is important to
keep in mind that (\ref{a.2}) is only valid when the first time
argument $x^0$ in (\ref{a.1}) is on the vertical part of the contour
while the second one is at $x^0=0$. If both arguments of the fields
are located on the vertical part according to (\ref{a.1}) one has to
use the time ordering along the imaginary time axis.

Since the order of the operators under the trace in (\ref{a.1}) can be
changed cyclically the real-time Wightman functions $G^{+-}$ and
$G^{-+}$ are related through
\begin{equation}
\label{a.3}
G^{+-}(x^0-\ii \beta,\vec{x})=G^{-+}(x^0,\vec{x}),
\end{equation}
where $x^0$ is a real-time argument on the contour and on the right
hand side one has to understand the analytic continuation. One should
keep in mind that only the Wightman functions are analytically
continueable, not the time ordered or anti-time ordered of the upper
or lower real-time branch, since they contain step functions from the
time ordering operator $\mathcal{T}_{\mathcal{C}}$. The same holds
true for the imaginary-time ordered Matsubara Green's function.

As in vacuum quantum field theory due to translation invariance it is
customary to use the energy-momentum representation of Green's
functions. For the real-time propagators we have the usual description
\begin{equation}
\label{a.4}
G^{ij}(x)=\feynint{p} \exp(-\ii p x) G^{ij}(p).
\end{equation}
We write down the formalism for arbitrary space-time dimensions since
all considerations do not depend on it and we need it to obtain well
defined non-renormalized quantities in the sense of dimensional
regularization. The periodic boundary condition (\ref{a.3}) translates
into the \emph{Kubo-Martin-Schwinger-condition} (KMS) for the Fourier
transformed Green's functions:
\begin{equation}
\label{a.5}
G^{-+}(p)=\exp(-\beta p_0) G^{+-}(p).
\end{equation}

The Matsubara Green's function is only defined for imaginary times
$-\ii \tau$ with $0\leq \tau \leq \beta$. Thus the momentum
representation with respect to the time component is a Fourier series
with period $\beta$ according to (\ref{a.3}) rather than a Fourier
integral:
\begin{equation}
\label{a.6}
G_M(x)=\frac{1}{\beta} \sum_{n=-\infty}^{\infty} \int
\frac{\d^{d-1} \vec{p}}{(2\pi)^{d-1}} \exp(-\ii p x) G_M(\ii
  p_0,\vec{p})|_{p^0=\omega_n} \text{ with } \omega_n=\frac{2
    \pi}{\beta} n. 
\end{equation}
Herein $x^0=-\ii \tau$ with $0\leq \tau \leq \beta$. Using the inverse
Fourier transform, Eq. (\ref{a.4}), and the KMS-condition (\ref{a.5})
we find the \emph{spectral representation}
\begin{equation}
\label{a.7}
G_M(\ii \omega_n,\vec{p})=\ii \int \frac{\d p_0}{2 \pi}
\frac{\rho(p_0,\vec{p})}{p_0-\ii \omega_n} \text { with } \rho(p)=\ii
[G^{+-}(p)-G^{-+}(p)].
\end{equation}
This shows that the Matsubara propagator is completely represented by
the real-time function $\rho$. With help of this we define the
analytically continued propagator by
\begin{equation}
\label{a.8}
G_c(k)=-\int \frac{\d p_0}{2 \pi}
\frac{\rho(p_0,\vec{k})}{p_0-k_{0}}. 
\end{equation}
It can contain singularities only on the real axis. Using the Fourier
transformation (\ref{a.4}) for the limits to the real axis
from above and below we obtain:
\begin{equation}
\label{a.9}
G_c(p_0 \pm \ii \eta,\vec{p}) = G_{R/A}(p), \; p_0 \in \R
\end{equation}
with the \emph{retarded and advanced} Green's functions
\begin{equation}
\label{a.10}
G_{R/A}(x)=\mp \ii \Theta(\pm t)
\erw{\comm{\op{\phi}(x)}{\op{\phi}(0)}}_{\beta}.
\end{equation}
From this we find immediately
\begin{equation}
\label{a.11}
G_R(x)=G_A^{*}(-x) \; \Rightarrow \; G_R(p)=G_A^*(p), \; \rho(p)=-2
\im G_R(p)=-\rho(-p) 
\end{equation}
and from (\ref{a.8}) and the analyticity of $G_c(p)$ in the
upper complex $p_0$-plane it follows that
\begin{equation}
\label{a.12}
\sigma(p_0)\rho(p) \geq 0.
\end{equation}
For later use we note the momentum space properties
\begin{equation}
\label{a.13}
G^{--}+G^{++}=G^{+-}+G^{-+}, \; G_R=G^{--}-G^{-+}, \;
G_A=G^{--}-G^{+-}, \; G_M=-\ii G_c(\ii \omega_n)
\end{equation}
which follow immediately from (\ref{a.1}), (\ref{a.7}), and
(\ref{a.10}).

We also make use of the expressions for the real-time Green's
functions in terms of the retarded Green's function, which follow
immediately from (\ref{a.7},\ref{a.11},\ref{a.15}):
\begin{align}
  \ii G^{--}(p) &= \ii G_R(p)+[\Theta(-p_0)+n(p_0)]
  \rho(|p_0|,\vec{p}), \label{gmm} \\
  \ii G^{++}(p) &=[\Theta(p_0)+n(p_0)] \rho(|p_0|,\vec{p})-\ii G_R(p),
  \label{gpp}\\ 
  \ii G^{-+}(p) &=[\Theta(-p_0)+n(p_0)] \rho(|p_0|,\vec{p}),
  \label{gmp} \\
  \ii G^{+-}(p) &=[\Theta(p_0)+n(p_0)] \rho(|p_0|,\vec{p}) \label{gpm}.
\end{align}
and the \emph{Bose-Einstein distribution defined as}
\begin{equation}
\label{a.15}
n(p_0)=\frac{1}{\exp(\beta|p_0|)-1}.
\end{equation}
All relations given above for the Green's functions $G$ directly apply
to any two-point function given by local field operators $\ii
h(x,y)=\erw{\mathcal{T}_{\mathcal{C}_{\R}}\op{H}(x)\op{H}(y)}$, e.g.,
the self-energy.

Real-time contour integrations and traces of translationally invariant
two-point functions
\begin{equation}\label{CR-Trace}
\begin{split}
C(x,y)&=\int_{\CR} \d z A(x,z)B(z,y),
\quad\quad x,y,z\in \CR \\
\TrR C&=\int_{\CR} \d x C(x,x)
\end{split}
\end{equation}
transcribe to 
\begin{equation}\label{CRp-Trace}
\begin{split}
C(p)^{ij}&=\sum_{kl}A(p)^{ik}\sigma_{kl}B(p)^{lj},
\quad\quad \sigma=\text{diag}(1,-1),\quad i,j,k,l\in \{-,+\}\\
\TrR C&=\sum_{ij}\feynint{p}C(p)^{ij}\sigma_{ij}
\end{split}
\end{equation}
in contour momentum-space representation.

We close this appendix by citing the formula for summation over the
Matsubara frequencies needed when calculating quantities related to
the vertical branch of the contour \cite{kap89}. In this paper we use
this to calculate the thermodynamical potential
\begin{equation}
\label{a.14}
\begin{split}
   \TrT h(p) & := \beta V \frac{1}{\ii \beta} \sum_{n=-\infty}^{\infty} \int
  \frac{\d^{d-1}\vec{p}}{(2\pi)^{d-1}} h(\ii
  \omega_n,\vec{p})\\
  & = -\beta V \feynint{p} \left [\frac{1}{2} +n(p_0) \right ] \left \{
    h[p_0+\ii \eta \sigma(p_0)] - h[p_0-\ii \eta \sigma(p_0)]
  \right \}\\
  & = \beta V \feynint{p}\left(h^{-+}+h^{+-}\right)
\end{split}  
\end{equation} 
In (\ref{a.14}) we have assumed that the function $h$ is analytic
below and above the real axis and that the trace exists.
Usually this is only the case for the regularized or the renormalized
functional traces. It is also clear that due to the exponential
damping from the Bose-Einstein distribution (\ref{a.15}) this part of
the integral has a superficial degree of divergence reduced by $1$
compared to the first part which is not damped by an $n$-factor.

Eq. (\ref{a.14}) shows that the thermodynamical potential can be
calculated from real-time quantities since the analytic continuation
of the Matsubara Green's function needed on the right hand side is
unique and can be obtained from the retarded Green's function as well
c.f. (\ref{a.9}). For a more detailed analysis of the analytic
properties see also \cite{lebel} and for the general case of Wigner
function representations in the non-equilibrium context
\cite{kv97}.

\begin{flushleft}

\end{flushleft}


\begin{thebibliography}{33}
\expandafter\ifx\csname natexlab\endcsname\relax\def\natexlab#1{#1}\fi
\expandafter\ifx\csname bibnamefont\endcsname\relax
  \def\bibnamefont#1{#1}\fi
\expandafter\ifx\csname bibfnamefont\endcsname\relax
  \def\bibfnamefont#1{#1}\fi
\expandafter\ifx\csname citenamefont\endcsname\relax
  \def\citenamefont#1{#1}\fi
\expandafter\ifx\csname url\endcsname\relax
  \def\url#1{\texttt{#1}}\fi
\expandafter\ifx\csname urlprefix\endcsname\relax\def\urlprefix{URL }\fi
\providecommand{\bibinfo}[2]{#2}
\providecommand{\eprint}[2][]{\url{#2}}

\bibitem[{\citenamefont{Luttinger and Ward}(1960)}]{lw60}
\bibinfo{author}{\bibfnamefont{J.}~\bibnamefont{Luttinger}} \bibnamefont{and}
  \bibinfo{author}{\bibfnamefont{J.}~\bibnamefont{Ward}},
  \bibinfo{journal}{Phys. Rev.} \textbf{\bibinfo{volume}{118}},
  \bibinfo{pages}{1417} (\bibinfo{year}{1960}),
  \urlprefix\url{http://prola.aps.org/abstract/PR/v118/i5/p1417_1}.

\bibitem[{\citenamefont{Lee and Yang}(1961)}]{leeyang61}
\bibinfo{author}{\bibfnamefont{T.~D.} \bibnamefont{Lee}} \bibnamefont{and}
  \bibinfo{author}{\bibfnamefont{C.~N.} \bibnamefont{Yang}},
  \bibinfo{journal}{Phys. Rev.} \textbf{\bibinfo{volume}{117}},
  \bibinfo{pages}{22} (\bibinfo{year}{1961}),
  \urlprefix\url{http://prola.aps.org/abstract/PR/v117/i1/p22_1}.

\bibitem[{\citenamefont{Baym and Kadanoff}(1961)}]{bk61}
\bibinfo{author}{\bibfnamefont{G.}~\bibnamefont{Baym}} \bibnamefont{and}
  \bibinfo{author}{\bibfnamefont{L.}~\bibnamefont{Kadanoff}},
  \bibinfo{journal}{Phys. Rev.} \textbf{\bibinfo{volume}{124}},
  \bibinfo{pages}{287} (\bibinfo{year}{1961}),
  \urlprefix\url{http://prola.aps.org/abstract/PR/v124/i2/p287_1}.

\bibitem[{\citenamefont{Baym}(1962)}]{baym62}
\bibinfo{author}{\bibfnamefont{G.}~\bibnamefont{Baym}}, \bibinfo{journal}{Phys.
  Rev.} \textbf{\bibinfo{volume}{127}}, \bibinfo{pages}{1391}
  (\bibinfo{year}{1962}),
  \urlprefix\url{http://prola.aps.org/abstract/PR/v127/i4/p1391_1}.

\bibitem[{\citenamefont{Cornwall et~al.}(1974)\citenamefont{Cornwall, Jackiw,
  and Tomboulis}}]{cjt74}
\bibinfo{author}{\bibfnamefont{M.}~\bibnamefont{Cornwall}},
  \bibinfo{author}{\bibfnamefont{R.}~\bibnamefont{Jackiw}}, \bibnamefont{and}
  \bibinfo{author}{\bibfnamefont{E.}~\bibnamefont{Tomboulis}},
  \bibinfo{journal}{Phys. Rev.} \textbf{\bibinfo{volume}{D10}},
  \bibinfo{pages}{2428} (\bibinfo{year}{1974}),
  \urlprefix\url{http://prola.aps.org/abstract/PRD/v10/i8/p2428_1}.

\bibitem[{\citenamefont{Schwinger}(1961)}]{Sch61}
\bibinfo{author}{\bibfnamefont{J.}~\bibnamefont{Schwinger}},
  \bibinfo{journal}{J. Math. Phys} \textbf{\bibinfo{volume}{2}},
  \bibinfo{pages}{407} (\bibinfo{year}{1961}).

\bibitem[{\citenamefont{Keldysh}(1964)}]{kel64}
\bibinfo{author}{\bibfnamefont{L.}~\bibnamefont{Keldysh}},
  \bibinfo{journal}{ZhETF} \textbf{\bibinfo{volume}{47}}, \bibinfo{pages}{1515}
  (\bibinfo{year}{1964}), \bibinfo{note}{[Sov. Phys JETP {\textbf{20}} 1965
  1018]}.

\bibitem[{\citenamefont{{S. A. Chin}}(1977)}]{chin77}
\bibinfo{author}{\bibnamefont{{S. A. Chin}}}, \bibinfo{journal}{Ann. Phys.}
  \textbf{\bibinfo{volume}{108}}, \bibinfo{pages}{301} (\bibinfo{year}{1977}).

\bibitem[{\citenamefont{Baym and Grinstein}(1977)}]{baymgrin}
\bibinfo{author}{\bibfnamefont{G.}~\bibnamefont{Baym}} \bibnamefont{and}
  \bibinfo{author}{\bibfnamefont{G.}~\bibnamefont{Grinstein}},
  \bibinfo{journal}{Phys. Rev.} \textbf{\bibinfo{volume}{D15}},
  \bibinfo{pages}{2897} (\bibinfo{year}{1977}),
  \urlprefix\url{http://prola.aps.org/abstract/PRD/v15/i10/p2897_1}.

\bibitem[{\citenamefont{Bielajew and Serot}(1984)}]{bielserot84}
\bibinfo{author}{\bibfnamefont{A.~F.} \bibnamefont{Bielajew}} \bibnamefont{and}
  \bibinfo{author}{\bibfnamefont{B.~D.} \bibnamefont{Serot}},
  \bibinfo{journal}{Annals of Physics} \textbf{\bibinfo{volume}{156}},
  \bibinfo{pages}{215} (\bibinfo{year}{1984}).

\bibitem[{\citenamefont{Ivanov et~al.}(1999)\citenamefont{Ivanov, Knoll, and
  Voskresensky}}]{kv97}
\bibinfo{author}{\bibfnamefont{Y.~B.} \bibnamefont{Ivanov}},
  \bibinfo{author}{\bibfnamefont{J.}~\bibnamefont{Knoll}}, \bibnamefont{and}
  \bibinfo{author}{\bibfnamefont{D.~N.} \bibnamefont{Voskresensky}},
  \bibinfo{journal}{Nucl. Phys.} \textbf{\bibinfo{volume}{A657}},
  \bibinfo{pages}{413} (\bibinfo{year}{1999}),
  \urlprefix\url{http://arXiv.org/abs/hep-ph/9807351}.

\bibitem[{\citenamefont{Knoll}(1999)}]{knoll98}
\bibinfo{author}{\bibfnamefont{J.}~\bibnamefont{Knoll}},
  \bibinfo{journal}{Prog. Part. Nucl. Phys.} \textbf{\bibinfo{volume}{42}},
  \bibinfo{pages}{177} (\bibinfo{year}{1999}),
  \urlprefix\url{http://arXiv.org/abs/nucl-th/9811099}.

\bibitem[{\citenamefont{Ivanov et~al.}(2000)\citenamefont{Ivanov, Knoll, and
  Voskresensky}}]{ikv99-2}
\bibinfo{author}{\bibfnamefont{Y.~B.} \bibnamefont{Ivanov}},
  \bibinfo{author}{\bibfnamefont{J.}~\bibnamefont{Knoll}}, \bibnamefont{and}
  \bibinfo{author}{\bibfnamefont{D.~N.} \bibnamefont{Voskresensky}},
  \bibinfo{journal}{Nucl. Phys.} \textbf{\bibinfo{volume}{A672}},
  \bibinfo{pages}{313} (\bibinfo{year}{2000}),
  \urlprefix\url{http://arXiv.org/abs/nucl-th/9905028}.

\bibitem[{\citenamefont{Leupold}(2000)}]{Leupold99}
\bibinfo{author}{\bibfnamefont{S.}~\bibnamefont{Leupold}},
  \bibinfo{journal}{Nucl. Phys.} \textbf{\bibinfo{volume}{A672}},
  \bibinfo{pages}{475} (\bibinfo{year}{2000}).

\bibitem[{\citenamefont{Kadanoff and Baym}(1961)}]{kb61}
\bibinfo{author}{\bibfnamefont{L.}~\bibnamefont{Kadanoff}} \bibnamefont{and}
  \bibinfo{author}{\bibfnamefont{G.}~\bibnamefont{Baym}},
  \emph{\bibinfo{title}{Quantum Statistical Mechanics}}, {A} {Lecture} {Note}
  and {Preprint} {Series} (\bibinfo{publisher}{The Benjamin/Cummings Publishing
  Company}, \bibinfo{year}{1961}).

\bibitem[{\citenamefont{Knoll and Voskresensky}(2001)}]{KIV01}
\bibinfo{author}{\bibfnamefont{Y.~I.} \bibnamefont{Knoll}, \bibfnamefont{J.}}
  \bibnamefont{and}
  \bibinfo{author}{\bibfnamefont{D.}~\bibnamefont{Voskresensky}},
  \bibinfo{journal}{Annals of Physics}  (\bibinfo{year}{2001}),
  \bibinfo{note}{in press},
  \urlprefix\url{http://arXiv.org/abs/nucl-th/0102044}.

\bibitem[{\citenamefont{van Hees and Knoll}(2001{\natexlab{a}})}]{vHK2001}
\bibinfo{author}{\bibfnamefont{H.}~\bibnamefont{van Hees}} \bibnamefont{and}
  \bibinfo{author}{\bibfnamefont{J.}~\bibnamefont{Knoll}},
  \bibinfo{journal}{Nucl. Phys.} \textbf{\bibinfo{volume}{A683}},
  \bibinfo{pages}{369} (\bibinfo{year}{2001}{\natexlab{a}}),
  \urlprefix\url{http://arXiv.org/abs/hep-ph/0007070}.

\bibitem[{\citenamefont{Blaizot et~al.}(2001)\citenamefont{Blaizot, Iancu, and
  Rebhan}}]{ianc00}
\bibinfo{author}{\bibfnamefont{J.~P.} \bibnamefont{Blaizot}},
  \bibinfo{author}{\bibfnamefont{E.}~\bibnamefont{Iancu}}, \bibnamefont{and}
  \bibinfo{author}{\bibfnamefont{A.}~\bibnamefont{Rebhan}},
  \bibinfo{journal}{Phys. Rev.} \textbf{\bibinfo{volume}{D63}},
  \bibinfo{pages}{065003} (\bibinfo{year}{2001}),
  \urlprefix\url{http://arXiv.org/abs/hep-ph/0005003}.

\bibitem[{\citenamefont{Peshier}(2001)}]{pesh00}
\bibinfo{author}{\bibfnamefont{A.}~\bibnamefont{Peshier}},
  \bibinfo{journal}{Phys. Rev.} \textbf{\bibinfo{volume}{D63}},
  \bibinfo{pages}{105004} (\bibinfo{year}{2001}),
  \urlprefix\url{http://arXiv.org/abs/hep-ph/0011250}.

\bibitem[{\citenamefont{Weinberg}(1960)}]{wein60}
\bibinfo{author}{\bibfnamefont{S.}~\bibnamefont{Weinberg}},
  \bibinfo{journal}{Phys. Rev.} \textbf{\bibinfo{volume}{118}},
  \bibinfo{pages}{838} (\bibinfo{year}{1960}),
  \urlprefix\url{http://prola.aps.org/abstract/PR/v118/i3/p838_1}.

\bibitem[{\citenamefont{Bogoliubov and Parasiuk}(1957)}]{bp57}
\bibinfo{author}{\bibfnamefont{N.~N.} \bibnamefont{Bogoliubov}}
  \bibnamefont{and} \bibinfo{author}{\bibfnamefont{O.~S.}
  \bibnamefont{Parasiuk}}, \bibinfo{journal}{Acta Math.}
  \textbf{\bibinfo{volume}{97}}, \bibinfo{pages}{227} (\bibinfo{year}{1957}).

\bibitem[{\citenamefont{Zimmermann}(1969)}]{Zim69}
\bibinfo{author}{\bibfnamefont{W.}~\bibnamefont{Zimmermann}},
  \bibinfo{journal}{Commun. Math. Phys.} \textbf{\bibinfo{volume}{15}},
  \bibinfo{pages}{208} (\bibinfo{year}{1969}).

\bibitem[{\citenamefont{Collins}(1986)}]{col86}
\bibinfo{author}{\bibfnamefont{J.~C.} \bibnamefont{Collins}},
  \emph{\bibinfo{title}{Renormalization}} (\bibinfo{publisher}{Cambridge
  University Press}, \bibinfo{address}{Cambridge, New York, Melbourne},
  \bibinfo{year}{1986}).

\bibitem[{\citenamefont{Kapusta}(1989)}]{kap89}
\bibinfo{author}{\bibfnamefont{J.~I.} \bibnamefont{Kapusta}},
  \emph{\bibinfo{title}{{Finite}-{Temperature} {Field} {Theory}}}
  (\bibinfo{publisher}{Cambridge University Press},
  \bibinfo{address}{Cambridge, New York, Melbourne}, \bibinfo{year}{1989}).

\bibitem[{\citenamefont{LeBellac}(1996)}]{lebel}
\bibinfo{author}{\bibfnamefont{M.}~\bibnamefont{LeBellac}},
  \emph{\bibinfo{title}{Thermal Field Theory}} (\bibinfo{publisher}{Cambridge
  University Press}, \bibinfo{address}{Cambridge, New York, Melbourne},
  \bibinfo{year}{1996}).

\bibitem[{\citenamefont{van Hees and
  Knoll}(2001{\natexlab{b}})}]{vHK2001-Ren-II}
\bibinfo{author}{\bibfnamefont{H.}~\bibnamefont{van Hees}} \bibnamefont{and}
  \bibinfo{author}{\bibfnamefont{J.}~\bibnamefont{Knoll}},
  \bibinfo{journal}{Phys. Rev. D}  (\bibinfo{year}{2001}{\natexlab{b}}),
  \bibinfo{note}{to be submitted}.

\bibitem[{\citenamefont{Landsmann and van Weert}(1987)}]{lvw87}
\bibinfo{author}{\bibfnamefont{N.~P.} \bibnamefont{Landsmann}}
  \bibnamefont{and} \bibinfo{author}{\bibfnamefont{C.~G.} \bibnamefont{van
  Weert}}, \bibinfo{journal}{Physics Reports} \textbf{\bibinfo{volume}{145}},
  \bibinfo{pages}{141} (\bibinfo{year}{1987}).

\bibitem[{\citenamefont{Gelis}(1996)}]{gel94}
\bibinfo{author}{\bibfnamefont{F.}~\bibnamefont{Gelis}}, \bibinfo{journal}{Z.
  Phys.} \textbf{\bibinfo{volume}{C70}}, \bibinfo{pages}{321}
  (\bibinfo{year}{1996}), \urlprefix\url{http://arXiv.org/abs/hep-ph/9412347}.

\bibitem[{\citenamefont{Itzykson and Zuber}(1980)}]{itz80}
\bibinfo{author}{\bibfnamefont{C.}~\bibnamefont{Itzykson}} \bibnamefont{and}
  \bibinfo{author}{\bibfnamefont{J.-B.} \bibnamefont{Zuber}},
  \emph{\bibinfo{title}{{Quantum} {Field} {Theory}}}
  (\bibinfo{publisher}{McGraw-Hill Book Company}, \bibinfo{address}{New York},
  \bibinfo{year}{1980}).

\bibitem[{\citenamefont{Zimmermann}(1970)}]{zim70}
\bibinfo{author}{\bibfnamefont{W.}~\bibnamefont{Zimmermann}},
  \bibinfo{journal}{Lectures on Elementary Particles and Quantum Field Theory.
  M.I.T. Press} p. \bibinfo{pages}{397} (\bibinfo{year}{1970}).

\bibitem[{\citenamefont{Danielewicz}(1984)}]{dan84}
\bibinfo{author}{\bibfnamefont{P.}~\bibnamefont{Danielewicz}},
  \bibinfo{journal}{Annals of Physics} \textbf{\bibinfo{volume}{152}},
  \bibinfo{pages}{239} (\bibinfo{year}{1984}).

\bibitem[{\citenamefont{van Hees}(2000)}]{vH-Ph-D00}
\bibinfo{author}{\bibfnamefont{H.}~\bibnamefont{van Hees}}, Ph.D. thesis,
  \bibinfo{school}{TU Darmstadt} (\bibinfo{year}{2000}),
  \urlprefix\url{http://elib.tu-darmstadt.de/diss/000082/}.

\bibitem[{\citenamefont{van Hees and
  Knoll}(2001{\natexlab{c}})}]{vHK2001-Ren-III}
\bibinfo{author}{\bibfnamefont{H.}~\bibnamefont{van Hees}} \bibnamefont{and}
  \bibinfo{author}{\bibfnamefont{J.}~\bibnamefont{Knoll}},
  \bibinfo{journal}{Phys. Rev. D}  (\bibinfo{year}{2001}{\natexlab{c}}),
  \bibinfo{note}{in preparation}.

\end{thebibliography}
\end{document}